\documentclass[11pt]{article}
\usepackage[a4paper,margin=1in]{geometry}
\usepackage{amsmath,amssymb,amsthm,booktabs,microtype}
\usepackage{xcolor,tikz,algorithm,algpseudocode,placeins}
\usepackage{authblk}
\definecolor{inkblue}{HTML}{215878}
\definecolor{ochre}{HTML}{99481A}
\definecolor{guide}{HTML}{637481}
\usepackage[font=small,labelfont=bf]{caption}
\usepackage[colorlinks=true,allcolors=blue!50!black]{hyperref}
\hypersetup{pdftitle={Online Network Design with Batch Processing},pdfauthor={Tianhang Lu, Runtian Ren, Shengcai Liu},pdfkeywords={online algorithms, TCP acknowledgment, parallel batch scheduling, flow time}}
\newtheorem{theorem}{Theorem}[section]
\newtheorem{lemma}[theorem]{Lemma}
\newtheorem{proposition}[theorem]{Proposition}
\newtheorem{corollary}[theorem]{Corollary}
\theoremstyle{definition}

\newcommand{\OPT}{\operatorname{OPT}}
\newcommand{\ALG}{\operatorname{ALG}}
\newcommand{\AtTime}{\Statex\textit{At each wake-up time \(t\), in order:}}
\title{Online Network Design with Batch Processing}
\author{Tianhang Lu, Runtian Ren, Shengcai Liu}
\affil{
Guangdong Provincial Key Laboratory of Brain-Inspired Intelligent Computation,\\
Department of Computer Science and Engineering,\\
Southern University of Science and Technology, Shenzhen 518055, China\\
liusc3@sustech.edu.cn
}
\date{}
\begin{document}
\maketitle
\begin{abstract}
An online service chooses which waiting jobs to start together. Each
finishes after its own processing time, and the longest determines the
next possible launch. We charge network cost plus either total flow or
the sum of per-batch maximum flows. For TCP acknowledgment with a fixed fee per
launch, the optimal deterministic total-flow ratio is \(1+\rho\), where
\(\rho\) is the processing-time spread. For batch maxima, a union of
intervals on the release-time axis gives the exact offline cost when
batches may overlap. It yields an offline \(2\)-approximation and pays
for an online timer of ratio five. The optimal deterministic ratio is
two at equal lengths and lies in \([2+2\sqrt2,5]\) at unrestricted spread.

How often can later batches incur the same blocking delay? Rooted
purchases can be combined, giving a polynomial-time \(3+2\sqrt2\)
guarantee for batch maxima on undirected Steiner networks. A mobile
server can concatenate walks in order, transferring an instantaneous
maximum-wait ratio \(\alpha\) to a batch-maximum flow ratio \(3\alpha+6\).
With \(k\) ephemeral facilities sharing one processor and one site per
launch, separate openings can each incur the wait. The optimal
deterministic batch-maximum ratio is then \(\Theta(k)\) at unrestricted
spread. The upper bound uses exact assignments, computable in polynomial
time for fixed \(k\).
\end{abstract}
\section{Introduction}
One client wants an image preview, another a fine detail. With
progressive encoding, each can read a prefix of the same
bitstream~\cite{progressive}. A single multicast serves both~\cite{multicast}.
Suppose the server has one outgoing stream and each transmission starts
at the beginning. A client arriving just after launch must wait for
the stream to become free. Admitting the client who wants more detail
extends that wait.

Machine-learning systems also make joint launch and execution decisions.
ED-Batch learns batching policies for dynamic neural
networks~\cite{chen2023edbatch}, while Packrat configures inference batch
sizes, threads, and model instances to reduce
latency~\cite{bhardwaj2025packrat}. These systems motivate the tradeoff
between grouping work and committing execution resources. We isolate
known job lengths and a shared processor with nonoverlapping launches;
the dependency, memory, and parallel-execution constraints of those
systems are outside our model.

Classical TCP acknowledgment charges one fee for any number of packets
acknowledged together~\cite{tcp}. We keep the fee and give each job a
processing length \(p_j\) revealed at arrival. Jobs admitted at \(s\)
finish at their individual times \(s+p_j\). The next launch must wait
for the longest admitted job. This is the unbounded \emph{drop-line}
batch machine~\cite{online-drop,dropline}.

\paragraph{Whom should a launch admit?}
In classical multilevel aggregation, a purchased subtree serves all
pending requests at its vertices~\cite{mla}. Consider one edge of cost
one. At time zero three unit jobs and a length-ten job arrive at its
far end. Three more unit jobs arrive at time one. Serving the first
three unit jobs at zero and the remaining four at one costs \(4+15=19\)
in total flow and fees. Under automatic admission, any launch before
time one blocks the second group until at least time ten. The cheapest
automatic schedule serves all seven at one and costs \(21\). Thus even
an offline scheduler benefits from choosing whom to admit. We give the
online algorithm and its offline comparator this choice, which we call \emph{elective service}.

\paragraph{How often is a blocked interval charged?}
Suppose two groups arrive while the processor is busy. Under total flow,
every job incurs its own wait. Under the sum of per-batch maximum flows,
the two groups incur one maximum if they leave together. If they leave
separately, both maxima can include the time spent waiting for the
processor. Each objective also charges the network actions performed.

The network determines which of these choices are available. On a
rooted distribution tree a transmission can buy the union of the paths
to both groups and pay once for the shared edges. In online service
with delay (OSD), reaching the first group changes the server's
position~\cite{osd}. Two postponed walks can still be executed at one
launch if the second starts where the first ends. Movement contributes
distance to the objective and takes place at launch. In facility
location with delay (FLD), opening a nearby site can save a long
connection~\cite{azar-touitou}. We consider ephemeral sites served by a
single shared processor, with one site per launch. Keeping the two
groups at different sites requires two launches. The second group can
therefore wait through both the original blockage and the first opening.

\paragraph{Our contributions.}
\begin{itemize}
\item \emph{Total flow.} Every job contributes \(p_j\) even in an
optimal schedule. This gives a common starting point for rooted
aggregation and elective OSD. An instantaneous total-wait algorithm of
ratio \(\alpha\) gives total-flow ratio \(\max\{\alpha,1+\rho\}\), where
\(\rho=p_{\max}/p_{\min}\). For fixed-fee TCP the optimal deterministic
ratio is exactly \(1+\rho\).
\item \emph{Batch maxima at one root.} For TCP, the optimal deterministic
ratio is two at equal lengths. For unrestricted lengths, every
deterministic algorithm has ratio at least \(2+2\sqrt2\), and our timer
has ratio exactly five. The interval formula behind its analysis also
gives an offline \(2\)-approximation, exact at equal lengths.
\item \emph{Combining purchases and walks.} For batch maxima, rooted
aggregation admits ratio \(3+2\sqrt2\) with exact monotone subadditive
purchases. The same bound is polynomial-time for rooted Steiner
purchases on undirected graphs. For elective OSD, an instantaneous
maximum-wait algorithm of ratio \(\alpha\) gives ratio \(3\alpha+6\).
Both bounds hold for arbitrary processing lengths.
\item \emph{Separate facility openings.} With \(k\) sites, we give a
\(k(3+2\sqrt2)\) batch-maximum bound using exact assignments,
computable in polynomial time for fixed \(k\). A lower bound of
\((k+1)/2\) on uniform metrics shows why the dependence on \(k\) is
necessary when the processing-time spread is unrestricted.
\end{itemize}

\paragraph{Our techniques.}
\begin{itemize}
\item \emph{A time interval for each TCP job.} Draw an interval of
length \(K+p_j\) ending at job \(j\)'s release, where \(K\) is the
opening fee. The length of the union equals the offline batch-maximum
cost if launches can overlap. Even when occupation forbids overlap,
its connected components pay for an online timer. A component can
have to pay for one batch that overruns its right endpoint.
\item \emph{Timing the final arrival.} Train jobs have \(p_j-r_j=1\),
so offline can wait and give every job the same flow in a single batch.
We release a final unit job once online has committed to its last launch
and the pooled offline batch has finished. The offline cost of this
last job is fixed. Online also pays for the time left in its last batch.
\item \emph{Executing postponed network decisions.} Run a request
algorithm on its own clock and record the actions it would take while
the processor is busy. For total flow, each job's own processing
provides a lower bound against which to charge the time it spends
waiting for those actions to be executed. For batch maxima this
per-job accounting can overcharge the optimum. For static purchases,
each duration can delay two rounds. Balancing that charge against the
cost of waiting gives a virtual price of \(c(B)+\sqrt2 P(B)\).
For walks, several online batches may select their longest job from the
same offline batch. We pack their occupied intervals into components
formed by offline flows and virtual waits. At most two batches per
component need a separate charge. At facilities each site can incur
the same delay, as a burst of short jobs at distinct sites demonstrates.
\end{itemize}

\paragraph{Related works.}
Dooly, Goldman, and Scott~\cite{tcp} analyze the two instantaneous TCP
waiting objectives. Frederiksen et al.~\cite{logp} model blocking
between transmissions by a prescribed minimum gap. Here the selected
jobs determine the gap. Tian et al.~\cite{online-drop} use the same drop-line machine
for maximum delivery completion time, and Gao et al.~\cite{dropline}
study offline regular completion-time objectives. A related objective
arises in batch delivery. Ji, He, and Cheng~\cite{batchdelivery} charge
weighted flow and delivery costs for jobs processed on one machine
and then delivered together. Here one network purchase launches the
jobs, each of which completes after its own processing time.

Christou et al.~\cite{lookahead2026} study total flow for online
batching with fixed processing duration, bounded capacity, and lookahead.
At equal lengths and zero opening fee, our one-root total-flow model
coincides with their uncapacitated case. Albers and van der
Heijden~\cite{active2026} combine active-time and flow-time costs for
preemptible jobs. Their active slots admit fresh work at subsequent
slot boundaries; our launches fix membership for the entire occupied
interval. These comparisons separate the information, occupation,
and fee assumptions behind the guarantees.

Azar and Touitou~\cite{azar-touitou} give a common framework for
aggregation, OSD, and facility location with delays or deadlines.
Their treatment makes explicit why these problems belong to the
same family even before processing is introduced. Work on per-batch
maximum waiting gives algorithms for elective walks~\cite{osdmax}
and static subadditive service costs~\cite{subadditive}. We use these
two results in different ways. A static purchase can be priced to reflect its future occupation.
For a mobile walk we keep the request algorithm's original movement
price and account for occupation after constructing the physical route.
Appendix~\ref{app:related} gives a more detailed comparison.

\paragraph{Organization.}
Section~\ref{sec:prelim} defines the two objectives and the
instantaneous comparison problems. Section~\ref{sec:tcp-main} gives
the TCP bounds. Its offline interval formula
(Section~\ref{sub:tcp-offline}) supplies the comparison for the online timer.
Sections~\ref{sec:aggregation-main}
and~\ref{sec:osd-main} each treat total flow first and batch maxima
second. Section~\ref{sec:facility-main} develops the common round
construction for facilities. Section~\ref{sec:conclusion} discusses
open questions. The appendices contain the full proofs and TCP
pseudocode.

\section{Preliminaries}\label{sec:prelim}
\paragraph{Jobs, actions, and clocks.}
Let \(J\) be a finite set of labelled jobs. Job \(j\) reveals its release
time \(r_j\ge0\), its processing time \(p_j>0\), and its location or
service type when it arrives. For a nonempty set \(B\), write
\(P(B)=\max_{j\in B}p_j\), with \(P(\varnothing)=0\). When \(J\ne\varnothing\), write
\(p_{\min}=\min_j p_j\), \(p_{\max}=\max_j p_j\), and
\(\rho=p_{\max}/p_{\min}\). Jobs at the same location or with identical
data remain distinct.

A single initially idle resource controls launches. At time \(s\), if
free, it may choose a feasible network action \(a\) and a nonempty set
\(B\) of released, unstarted jobs that the action can serve. The action
is executed instantaneously, all jobs in \(B\) start at \(s\), and the
resource is occupied on \([s,s+P(B))\). No additions or preemptions
are allowed. Job \(j\in B\) finishes at \(C_j=s+p_j\), with flow
\(F_j=C_j-r_j\). The network model specifies the spatial charge \(c(a)\)
and, for mobile service, the terminal position. A schedule launches
each job exactly once. Under elective service, both the algorithm and the offline comparator
choose the admitted set. Other covered jobs remain pending, even at
a vertex of the purchased subtree or a location visited by the walk.

\paragraph{The two objectives.}
Let \(\mathcal B\) be the schedule's nonempty batches and \(N\) its
total network cost, including any empty moves of the mobile server.
Write \(W(B,s)=s-\min_{j\in B}r_j\) for the oldest wait at launch and
\(M(B,s)=\max_{j\in B}(s-r_j+p_j)\) for maximum flow. We minimize
\begin{align}
  \operatorname{cost}_{\rm sum}&=N+\sum_{j\in J} F_j,\label{eq:sum}\\
  \operatorname{cost}_{\rm batch}&=N+\sum_{B\in\mathcal B}M(B,s_B).
      \label{eq:batch}
\end{align}
For TCP, \(c(a)=K>0\) per launch. In the other models \(c(a)\) measures
the purchased subtree, walk, or facility service. We will use
\(M(B,s)\ge P(B)\) and \(M(B,s)\ge W(B,s)+P(B)/\rho\); the second
inequality follows by considering an oldest job. We write
\(x_+=\max\{x,0\}\).

\paragraph{Information and comparison schedules.}
Arrivals at a common time are revealed together before any launch. A
batch ending then frees the resource. The input length, end of input,
and bounds on future lengths are unannounced unless specified. Every
job must eventually finish on every finite input. For either objective,
\(\OPT\) is the cost of a best offline schedule under the same rules.
A strict deterministic ratio \(R\) means \(\ALG(I)\le R\OPT(I)\)
on every input. The empty input costs zero.

The corresponding \emph{instantaneous request problem} removes
occupation and charges waiting only up to the action time. We denote its
optimum by \(\OPT_0\), always using the same network and delay
aggregation as specified. Some reductions explicitly change the virtual
service cost. An exact static batch oracle returns both the minimum cost
and a feasible action (or assignment) realizing it. An oracle guarantee
has a polynomial-time implementation when the needed oracle calls do.

\paragraph{An instantaneous batching fact.}
We use the following deterministic result of Lu, Ren, and
Liu~\cite{subadditive}. For a nonnegative, normalized, monotone
subadditive batch cost \(h\) with exact realization, serve the entire
pending set \(A\) when its oldest wait reaches \(h(A)\). This rule is
\(2\)-competitive for total purchase cost plus the sum of per-batch
maximum waits. For costs depending on a job's individual connection or
length, take each labelled job as a distinct type. On any finite input
the theorem applies on its finite set of labels. Its threshold queries
only the current pending set, so enlarging the type set by future
labels does not change any earlier decision.

\section{TCP acknowledgment with batch processing}\label{sec:tcp-main}
At one root the only spatial decision is whether to pay \(K\) again.
This isolates the cost of a launch that prevents service of later jobs.
The two objectives give sharply different answers (Table~\ref{tab:results}).

\subsection{Total flow time}
\begin{theorem}[TCP, total flow]\label{thm:tcp-front-sum}
For \eqref{eq:sum} the optimal deterministic ratio at each fixed
spread \(\rho\) is \(1+\rho\). The same algorithm works for every spread.
\end{theorem}
Run a virtual acknowledgment whenever the sum of the pending requests'
waits reaches \(K\). Buffer these decisions during physical processing
and launch their union when the root becomes free. A job waits at most
one more batch duration after its virtual acknowledgment. Every offline
schedule already pays the job's own processing time in flow. The ratio
between these two lengths is at most \(\rho\). Conversely, a burst of
unit jobs released just after the longest job starts makes nearly all
flow come from jobs that have to wait through this commitment. More
precisely, if the running job has length \(\rho\), each new job pays
nearly \(\rho+1\) in flow. Offline can hold the initial jobs until the
burst arrives and start everyone together. Each new job then pays one,
and increasing the size of the burst makes the initial jobs' costs and
the opening fee negligible. For a batch maximum the size of this burst
would buy the adversary nothing more once one unit job was present.
Appendix~\ref{sec:total} gives the algorithm and matching bounds.

\subsection{Offline batch maxima and the interval formula}\label{sub:tcp-offline}
Suppose all releases and lengths are known. How should we group the
jobs if two batches may run at the same time? Each batch still pays
the fee \(K\), and job \(j\) launched at \(s\) still completes at
\(s+p_j\). Denote the optimum for this relaxation of~\eqref{eq:batch}
by \(\OPT_{\rm overlap}\). The following formula gives both an optimal
partition for the relaxation and a way to schedule it on one processor.

\begin{theorem}[Offline batch maxima]\label{thm:tcp-offline-front}
For \(I_j=[r_j-K-p_j,r_j]\) and \(U=\bigcup_j I_j\), we have
\(\OPT_{\rm overlap}=|U|\). On one processor, an \(O(n\log n)\)-time
algorithm constructs a schedule of cost at most \(2|U|\le2\OPT\).
If all jobs have the same length, launching each component's jobs at
its right endpoint attains \(|U|=\OPT\).
\end{theorem}

\paragraph{Why the union gives the exact price.}
The right endpoint of \(I_j\) is the first time job \(j\) can start.
Its left endpoint is chosen so that the distance from there to any
legal launch \(s\) is exactly \(K+s-r_j+p_j\). For an entire batch
\(B\), taking its largest individual flow therefore gives the identity
\[
    K+M(B,s)=s-\min_{j\in B}(r_j-K-p_j).
\]
The interval from that minimum to \(s\) covers every \(I_j\) in the
batch. Summing the lengths of these covers over any schedule pays at
least \(|U|\). Conversely, take one connected component \([L,R]\) of
\(U\). All its jobs have arrived by \(R\), and one of their intervals
starts at \(L\). Launching them together at \(R\) costs exactly \(R-L\).
One such launch per component attains the lower bound when batches may
overlap.

\paragraph{When do these launches already fit?}
With common processing time \(p\), every interval has width \(K+p\).
If one component ends at \(R\) and the next has first release \(a\),
then \(a-(K+p)>R\), since otherwise their intervals would touch.
The batch launched at \(R\) thus finishes at \(R+p\) before even the
first release of the next component. With unequal lengths, a short
interval can start a new component while an earlier long job is still
running. For example, take \(K=1\) and three jobs with
\((r_j,p_j)=(0,8),(3,1),(6,1)\). Their intervals
\([-9,0],[1,3],[4,6]\) are disjoint. The relaxed optimum pays
\(9+2+2=13\), but the long job launched at zero occupies the processor
at both \(3\) and \(6\). Launching the two waiting jobs together at
\(8\) costs \(1+(9-3)=7\), for a feasible total of \(9+7=16\).

\paragraph{Why restoring feasibility costs at most a factor of two.}
Visit the component right endpoints in order. Launch a component's
jobs at its endpoint if the processor is free. Otherwise buffer them.
As soon as a batch finishes, launch all buffered components together,
including any due at that instant. This rule makes a component wait
through at most the remainder of one physical batch, even if many
endpoints fall during that occupation.

For physical batch \(i\), let \(H_i\) be the largest length among
the components merged into it, and let \(P_i\) be its longest
processing time, with \(P_0=0\). At its component endpoint, every
job's individual flow plus one fee was at most \(H_i\). Delaying
its start adds at most \(P_{i-1}\) to its flow. The merged batch
pays one fee and takes one maximum, so its cost is at most
\(H_i+P_{i-1}\). Also \(P_i\le H_i\), since its longest job has
an interval of width \(K+P_i\) inside one of these components.
Every component occurs in exactly one merged group. Consequently
\[
  \sum_i (H_i+P_{i-1})\le 2\sum_i H_i\le 2|U|.
\]
One copy of the union pays for the component batches themselves.
The second pays for the blocking each physical batch can pass to its
successor. Sorting the interval endpoints takes \(O(n\log n)\) time,
followed by a linear scan. This factor is tight for the repair rule.
With \(K=1\) and jobs \((0,P),(3,1)\), \(P>3\), the rule pays
\(2P\), whereas starting both jobs at time three costs \(P+4\).

\begin{proposition}[Few processing lengths]
For \(d\) distinct processing lengths, either offline TCP objective
can be minimized in \(n^{O(d)}\) time.
\end{proposition}
Within a length class, an optimum can serve jobs in release order.
For batch maxima, move a later arrival into the batch of an earlier
arrival of the same length whenever their order is inverted: the
destination batch already contains a job whose flow dominates its flow.
For total flow, swap the two jobs. This leaves one served prefix per
length class to record, together with the previous finishing time.
Appendix~\ref{app:offline-types} bounds the time grid and gives the
recurrence. The equal-length total-flow algorithm takes \(O(n^4)\) time.

\subsection{Batch maxima and the pending timer}
\begin{figure}[htbp]
\centering
\begin{tikzpicture}[x=.83cm,y=.58cm,font=\small]
 \foreach \t in {0,2,6,7,9,11,12} {
   \draw[black!18] (\t,0)--(\t,4.65);
   \node[below] at (\t,-.06) {\scriptsize $\t$};}
 \draw[->] (-.2,0)--(12.8,0) node[right] {\scriptsize time};
 \foreach \row/\job/\release/\start/\finish in {3/A/0/6/7,2/B/1.5/6/11,1/C/7/11/12} {
   \node[left] at (-.28,\row) {Job \(\job\)};
   \draw[guide,thick,dashed] (\release,\row)--(\start,\row);
   \filldraw[fill=white,draw=guide,thick] (\release,\row) circle (2.1pt);
   \fill[inkblue] (\start,\row-.15) rectangle (\finish,\row+.15);}
 \node[left] at (-.28,4) {Root};
 \filldraw[fill=inkblue!12,draw=inkblue] (6,3.70) rectangle (11,4.3);
 \filldraw[fill=ochre!12,draw=ochre] (11,3.70) rectangle (12,4.3);
 \node at (8.5,4) {\scriptsize $\{A,B\}$};
 \node at (11.5,4) {\scriptsize $\{C\}$};
\end{tikzpicture}
\caption{A timer postponed by an arrival and a timer blocked by the root.
Circles mark releases, dashes waiting, and solid bars processing.
With \(K=1\), job \(A\) of length 1 sets a timer for 2. At time \(1.5\),
job \(B\) of length 5 moves it to 6. Job \(A\) finishes at 7, when
unit job \(C\) arrives. Its timer expires at 9, but its batch can start
only at 11. The three flow times are \(7,9.5,5\), giving total-flow
cost \(23.5\) and batch-maximum cost \(16.5\), fees included.}
\label{fig:timeline}
\end{figure}
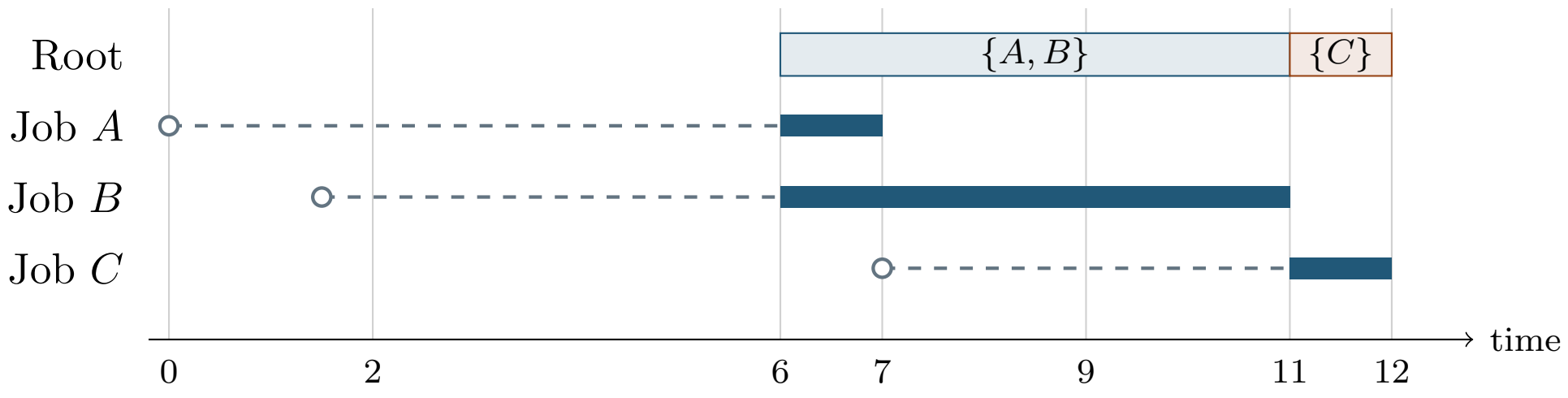

\begin{theorem}[TCP, sum of batch maxima]\label{thm:tcp-front-max}
For \eqref{eq:batch} the optimal deterministic ratio is two at equal
lengths. With unrestricted unannounced lengths it is in \([c_*,5]\),
where \(c_*=2+2\sqrt2=4.8284271\ldots\). The pending timer has exact ratio five.
\end{theorem}
Let \(b\) be the oldest pending release. Launch all pending jobs as soon
as the root is free and the time is at least \(b+K+P(B)\). A long job
arriving before launch can postpone the timer; an earlier batch can
block a timer that has already expired. Figure~\ref{fig:timeline}
exhibits both effects. At a common length \(p\) the timer always gets
its full \(K+p\) units of waiting, by which time the previous
occupation has ended. Its batch costs exactly \(2(K+p)\). As soon as
lengths differ, even a single short job can wait well beyond its timer.
We first give an accounting that pays for this excess on every input,
then an adversary that makes it unavoidable.
\begin{table}[htbp]
\centering
\caption{Deterministic competitive ratios for TCP-BP. The last row allows arbitrarily large, unannounced processing times.}\label{tab:results}
\small
\begin{tabular}{@{}llcc@{}}
\toprule
Objective & Processing times & Lower bound & Upper bound\\
\midrule
Total flow & Spread \(\rho\ge1\) & \(1+\rho\) & \(1+\rho\)\\
Per-batch maximum & All equal & \(2\) & \(2\)\\
Per-batch maximum & Fixed spread \(\rho>1\)
 & \(1+2/\sqrt{3+1/\rho}\) & \(\min\{5,2\rho\}\)\\
Per-batch maximum & Unrestricted & \(2+2\sqrt2\) & \(5\)\\
\bottomrule
\end{tabular}
\end{table}

\paragraph{Why five copies of the interval union suffice.}
The timer chooses its batches before the components of \(U\) are known.
We bound its cost against the offline value \(|U|\) from
Section~\ref{sub:tcp-offline} by identifying which component pays for each wait.
For timer batch \(i\), write \(a_i\) for its oldest release, \(s_i\)
for its start, \(P_i\) for its longest length, and \(d_i=K+P_i\).
Its wait \(w_i=s_i-a_i\) is \(d_i+\delta_i\), where \(\delta_i\) is
the excess after the timer expires. The preceding batch cleared all
pending jobs when it started. Thus it can block the new oldest job
for at most \(P_{i-1}\) time units, of which the timer already allows
\(d_i\). Writing \(C_i\) for the fee plus maximum flow, we obtain
\[
       C_i\le w_i+d_i=2d_i+\delta_i\le2w_i,\qquad
       \delta_i\le(d_{i-1}-d_i)_+ \quad(i\ge2).
\]
The first batch has zero excess. Charging twice the wait of every
batch would still leave a problem: the last wait can run well past
the interval component containing the jobs that pay for it.

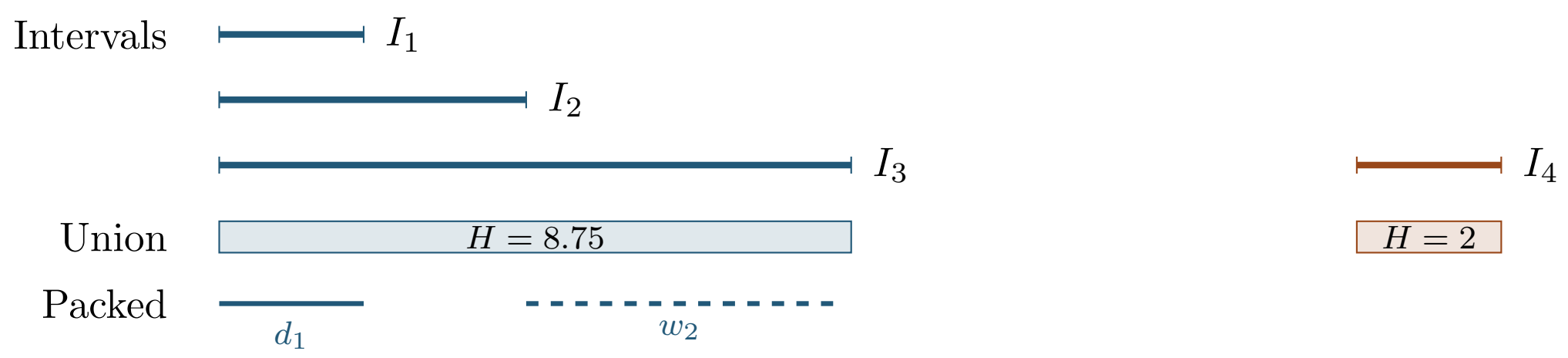
\begin{figure}[htbp]
\centering
\begin{tikzpicture}[x=.62cm,y=.56cm,font=\small]
 \node[left] at (-2.5,4) {Intervals};
 \foreach \t/\y/\name in {0/4/1,2.25/3/2,6.75/2/3}{
  \draw[inkblue,line width=1.6pt] (-2,\y)--(\t,\y);
  \draw[inkblue] (-2,\y-.13)--(-2,\y+.13) (\t,\y-.13)--(\t,\y+.13);
  \node[right] at (\t+.08,\y) {$I_{\name}$};}
 \draw[ochre,line width=1.6pt] (13.75,2)--(15.75,2);
 \draw[ochre] (13.75,1.87)--(13.75,2.13) (15.75,1.87)--(15.75,2.13);
 \node[right] at (15.83,2) {$I_4$};
 \node[left] at (-2.5,.9) {Union};
 \filldraw[fill=inkblue!14,draw=inkblue] (-2,.66) rectangle (6.75,1.14);
 \filldraw[fill=ochre!15,draw=ochre] (13.75,.66) rectangle (15.75,1.14);
 \node at (2.375,.9) {\scriptsize $H=8.75$};
 \node at (14.75,.9) {\scriptsize $H=2$};
 \node[left] at (-2.5,-.12) {Packed};
 \draw[inkblue,very thick] (-2,-.12)--(0,-.12) node[midway,below] {\scriptsize $d_1$};
 \draw[inkblue,very thick,dashed] (2.25,-.12)--(6.5,-.12) node[midway,below] {\scriptsize $w_2$};
\end{tikzpicture}
\caption{Interval accounting for \(K=1\) and jobs \((r,p)=(0,1),
(2.25,3.25),(6.75,7.75),(15.75,1)\). The timer starts its first
three batches at \(2,6.5,15.5\). The first backward interval and the
middle batch's wait fit disjointly in the left component. The last job's
timer expires at \(17.75\), but the third batch runs until \(23.25\).
Its excess wait of \(5.5\) is charged to the preceding component.}
\label{fig:packing}
\end{figure}

Assign each batch to the component containing the backward interval
of one of its longest jobs. These representatives arrive in batch
order because each launch clears the pending set. Fix one used
component \([L,R]\) of length \(H\) and number its batches locally
from \(1\) to \(k\). Three charges account for its cost.
\begin{enumerate}
\item \emph{First and interior batches.} The first representative's
interval of length \(d_1\) lies inside the component. Every interior
wait begins after that representative's release and ends before the
last representative's release. The waits are disjoint, since the next
batch's first job arrives after the preceding launch. Thus
\(d_1+\sum_{i=2}^{k-1}w_i\le H\) for \(k\ge2\). Charging
\(2d_1\) for the first batch and \(2w_i\) for the interior ones uses
at most \(2H\), leaving only the first batch's excess unpaid.
\item \emph{Last batch.} If \(k\ge2\), the preceding batch also belongs
to this component. Both \(d_{k-1}\) and \(d_k\) are at most \(H\).
Even if the last wait extends far beyond \(R\), its cost is at most
\(2d_k+(d_{k-1}-d_k)_+=d_k+\max\{d_{k-1},d_k\}\le2H\).
For a component with only one batch, \(C_1\le2H+\delta_1\) directly.
\item \emph{Excess passed to the next component.} The first batch of a
new used component can still be blocked by the last batch of the
previous used component. Charge this excess to that previous component:
its last \(d_i\) is at most its length. Only its immediate successor
can make this charge. Each component thus pays at most one further \(H\).
\end{enumerate}
In Figure~\ref{fig:packing}, the first interval and the middle wait
fit inside the left component. The short job on the right outwaits
its own timer by \(5.5\), more than its component's entire length of
\(2\). The component on the left pays the excess. Summing the three
charges gives \(\ALG\le(2+2+1)|U|\le5\OPT\).
Appendices~\ref{sec:max-upper} and~\ref{sec:fixed-spread} give the
full proof and the spread-sensitive bound
\(\min\{5,2(K+p_{\max})/(K+p_{\min})\}\) for the same timer.

\subsection{The cost of the last occupation}
To obtain a lower bound for \emph{every} deterministic algorithm,
consider creating one new job just after each launch. The algorithm
must then pay for each job in its own batch. This is expensive only if
offline can still combine the jobs cheaply. At a common launch time
\(s\), job \(j\)'s flow is \(s+(p_j-r_j)\). Set \(r_1=0\) and
\(p_j=r_j+1\) throughout the sequence. No matter how long it grows,
all jobs have exactly the same flow when launched together.

Fix \(0<\epsilon<1\), let \(T_i\) be the algorithm's launch of job
\(i\), and release the next long job at \(T_i+\epsilon\) whenever
the sequence continues. Write \(b_i=T_i+K+1\) for the singleton
batch's cost in this argument, and \(S_i=\sum_{j=1}^i b_j\). At any
launch \(T_i\), \(i\ge2\), the algorithm has committed to both
\(S_i\) in fees and flows and an occupation of length \(p_i\). Ending the input here isolates the prefix cost. One more job will
expose the remaining occupation on the same prefix.

\begin{itemize}
\item \emph{Stop.} Offline opens once at the final release
\(r_i=T_{i-1}+\epsilon\), paying \(b_{i-1}+\epsilon\).
This tests the accumulated singleton costs \(S_i\). The last job's
own flow is included, but no later job pays for being blocked by it.
\item \emph{Append a unit job when the comparator is free.}
The offline train launched at \(r_i\) finishes at \(r_i+p_i\).
Release the unit job at \(R_i=\max\{T_i+\epsilon,r_i+p_i\}\).
Offline starts it immediately and pays just \(K+1\).
Online still has the remaining occupation
\[
             T_i+p_i-R_i=\min\{p_i-\epsilon,T_i-r_i\}.
\]
If offline finishes first, the unit job experiences almost the full
online occupation. If offline finishes later, the two batches contain
the same longest job, and the residual is exactly the difference
between their launch times. This choice of release makes the
comparator's added cost constant in both cases.
\end{itemize}

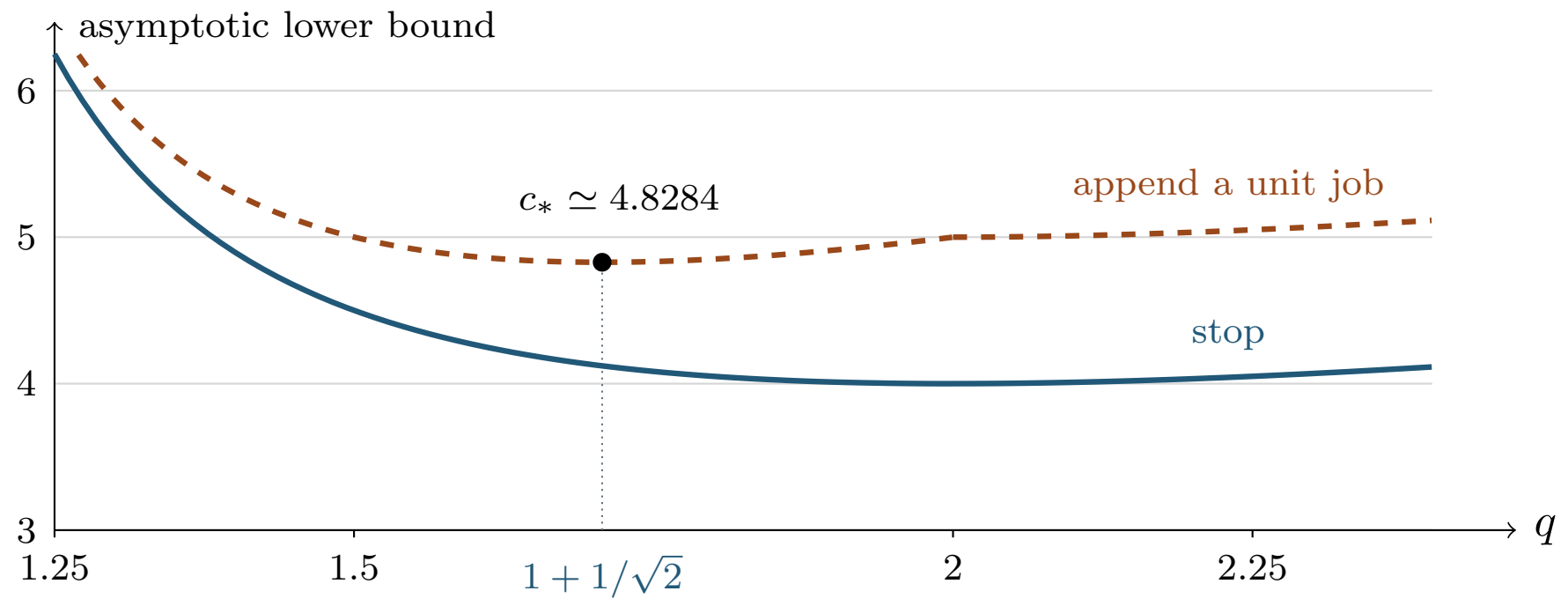
\begin{figure}[htbp]
\centering
\begin{tikzpicture}[x=9cm,y=1.10cm,font=\small]
\begin{scope}[shift={(-1.25,-3)}]
 \foreach \y in {3,4,5,6}{
   \draw[black!16] (1.25,\y)--(2.40,\y);
   \node[left] at (1.25,\y) {\scriptsize $\y$};}
 \draw[->] (1.25,3)--(2.47,3) node[right] {$q$};
 \draw[->] (1.25,3)--(1.25,6.47);
 \node[anchor=west] at (1.255,6.43) {\scriptsize asymptotic lower bound};
 \foreach \x in {1.25,1.5,2,2.25}{
   \draw (\x,3)--(\x,2.95) node[below] {\scriptsize $\x$};}
 \draw[guide,densely dotted] (1.707107,3)--(1.707107,4.828427);
 \node[below,inkblue] at (1.707107,2.96) {\scriptsize $1+1/\sqrt2$};
 \draw[inkblue,very thick,domain=1.25:2.4,samples=100,variable=\x]
     plot (\x,{\x*\x/(\x-1)});
 \draw[ochre,very thick,dashed,domain=1.27:2,samples=85,variable=\x]
     plot (\x,{\x*\x/(\x-1)+\x-1});
 \draw[ochre,very thick,dashed,domain=2:2.4,samples=45,variable=\x]
     plot (\x,{(\x*\x/(\x-1)+1)});
 \fill[black] (1.707107,4.828427) circle (2pt);
 \node[anchor=south east,inner sep=1pt] at (1.81,5.12) {\scriptsize $c_*\simeq4.8284$};
 \node[inkblue,above] at (2.23,4.11) {\scriptsize stop};
 \node[ochre,above] at (2.23,5.12) {\scriptsize append a unit job};
\end{scope}
\end{tikzpicture}
\caption{The prefix cost and the extra cost of blocking a unit job. For
\(q=\limsup b_i/b_{i-1}\), stopping gives \(f(q)\) (solid).
Releasing a unit job when the offline processor is free gives
\(f(q)+\min\{1,q-1\}\) (dashed). Its minimum is \(2+2\sqrt2\).
At the pending timer's growth rate \(q=2\), the bound is five.}
\label{fig:forks}
\end{figure}

\paragraph{What fixes the constant.}
Suppose first that the singleton costs \(b_i\) grow geometrically by
a factor \(q>1\). In units of \(b_{i-1}\), their accumulated cost
tends to \(f(q)=q+1+q^{-1}+\cdots=q^2/(q-1)\).
The input with the final unit job already incurs this entire prefix
cost. On that same input the last job adds a residual of asymptotic size \(\min\{1,q-1\}\), while the
offline cost remains \(b_{i-1}+O(1)\). For \(q\ge2\) this gives
\(f(q)+1\ge5\). For \(1<q\le2\), writing \(x=q-1\) gives
\[
       f(q)+q-1=2+2x+1/x\ge 2+2\sqrt2 .
\]
The minimum occurs at \(q=1+1/\sqrt2\)
(Figure~\ref{fig:forks}). Launching more often increases the
normalized prefix cost, whereas launching later leaves a larger
residual when the comparator finishes.

The adversary does not require geometric launches. Take
\(q=\limsup_i b_i/b_{i-1}\) and a subsequence approaching this
value. Every sufficiently late growth factor is at most \(q+\eta\)
for any fixed \(\eta>0\). Looking back along the prefix gives at
least the corresponding truncated geometric sum; letting \(\eta\)
decrease to zero and the number of terms grow yields \(f(q)\).
The residual bound applies at those same indices. For \(q=1\) or
\(q=\infty\) the prefix bound alone diverges. Hence every ratio
below \(2+2\sqrt2\) fails on a finite continuation.

\paragraph{Why the timer itself reaches five.}
On this sequence each pending set is a singleton. Its timer for job
\(i\) expires at \(r_i+K+p_i=2r_i+K+1\), by which time the previous
batch has finished. Hence \(b_i=2b_{i-1}+2\epsilon\): the timer
realizes exactly the limiting growth rate \(q=2\). Stopping accounts
for four units of normalized cost. The final unit job adds a fifth,
while offline serves it immediately for \(K+1\), negligible compared
with \(b_{i-1}\). The upper bound is therefore tight for this timer.
Whether another deterministic algorithm can improve on five remains
open. Appendix~\ref{sec:max-lower} gives the finite-prefix
inequalities and the minimization in full.

\paragraph{Which lower bounds carry over?}
An aggregation tree with a single positive-cost edge, or one facility
with opening fee \(K\) and colocated clients, realizes the TCP instance
exactly. Lower bounds from Table~\ref{tab:results} therefore hold for
these network families. A mobile server parked at one point pays zero
movement. The unrestricted train construction still works with \(K=0\):
the offline schedule and normalized prefix inequalities in
Appendix~\ref{sec:max-lower} remain valid with \(K+1=1\).
This yields the same lower bound \(c_*\) for the mobile batch-maximum
objective. The burst argument for total flow likewise allows zero fees.
\FloatBarrier

\section{Rooted aggregation}\label{sec:aggregation-main}
In multi-level aggregation (MLA), each job arrives at a vertex of a
weighted tree. A launch buys a subtree connecting the root to the jobs
selected for the batch~\cite{mla}. Let \(c(B)\) denote the weight of
the union of their root paths. On a rooted half-line this is the
distance to the farthest selected job, as in line aggregation with
delay (LAD). In joint replenishment (JRP) the common order fee is a
shared edge and each included item type adds a branch. Jobs at two
locations can therefore share a purchase even if their processing
times differ substantially.

Consider a tree with a trunk of cost \(u\) and two terminal branches of
cost \(v\) and \(w\). While the processor is busy, an instantaneous
algorithm might make one purchase reaching the first branch and later
another reaching the second. It pays \(2u+v+w\). Once the processor is
free, one launch reaches both for \(u+v+w\). If the longest jobs in
the two groups have lengths \(p\) and \(q\), the merged batch occupies
the root for \(\max\{p,q\}\). The spatial saving is immediate. Its
effect on delay depends on whether the objective counts all the jobs
that had to wait or only the slowest response in the merged batch.

\subsection{Total flow time}\label{sub:rooted-total}
Simulate an algorithm for the instantaneous request problem, giving it
the actual release times and locations. We call a job \emph{buffered}
once this algorithm has purchased service for it but the physical
processor has yet to launch it. The virtual algorithm can keep buying
subtrees while the processor is busy. At its next opportunity the
physical algorithm buys their union and starts the buffered jobs. A
newly arrived job that the virtual algorithm has not selected stays
pending, even if its location lies on this subtree.

\begin{theorem}[Total flow for rooted aggregation]\label{thm:rooted-front}
An instantaneous \(\alpha\)-competitive algorithm for network cost
plus total waiting yields a \(\max\{\alpha,1+\rho\}\)-competitive
algorithm for network cost plus total flow. The job algorithm needs
only unions of the supplied actions and does not need to know \(\rho\).
\end{theorem}

\paragraph{Proof sketch.}
Let \(a_j\) be the virtual service time of job \(j\), and let \(s_j\)
be its physical start. If the processor is idle at \(a_j\), the job
starts immediately. Otherwise it starts as soon as the current batch
ends. Even if many virtual purchases take place in the meantime,
none of them requires a separate physical launch ahead of \(j\).
Thus \(0\le s_j-a_j\le p_{\max}\). Each purchased virtual subtree
contributes to a single union, so the total spatial cost of these
unions is at most the virtual network cost \(N_V\).

Write \(P_\Sigma=\sum_jp_j\), and let \(G_V=\sum_j(a_j-r_j)\)
be virtual waiting. A physical job pays its virtual wait, the extra
time \(s_j-a_j\), and its own processing. Since
\(p_{\max}\le\rho p_j\) for every job, we obtain
\[
   \ALG_{\rm sum}\le N_V+G_V+(1+\rho)P_\Sigma.
\]
For the reverse comparison, take an optimal physical schedule. Keep
every subtree and its launch time, and serve its jobs instantaneously
at that time. This subtracts exactly \(P_\Sigma\) from the schedule's
cost, leaving a feasible request schedule. Thus
\(\OPT_0+P_\Sigma\le\OPT\). Applying the virtual guarantee gives
\[
   \ALG_{\rm sum}\le\alpha\OPT_0+(1+\rho)P_\Sigma
          \le\max\{\alpha,1+\rho\}\OPT.
\]
The two coefficients multiply different parts of the same offline
schedule. In particular, the buffering step introduces no additional
dependence on tree depth. Even a single edge can incur the factor
\(1+\rho\), by the short-job burst in Section~\ref{sec:tcp-main}.

\subsection{The sum of batch maximum flow times}\label{sub:rooted-max}
For batch maxima, \(m\) equal jobs of length \(p\) launched together
at release incur a single charge \(p\). The additive proof's budget
of \(\sum_j p_j\) is unavailable. At equal lengths there is an exact
way to include processing in a virtual purchase: charge \(c(B)+p\)
for every nonempty batch and use the instantaneous all-pending timer.
After a launch, even the next oldest job arrives later and then waits
at least \(p\) before its timer fires. Each virtual launch is feasible
at its prescribed time. Its virtual cost is exactly its physical cost.
Deleting occupation gives the reverse comparison for the optima.
The instantaneous factor two is therefore preserved.

\begin{theorem}[Batch maxima for rooted purchases]\label{thm:rooted-max}
LAD, JRP, and MLA have optimal deterministic ratio two at equal lengths.
For a nonnegative normalized monotone subadditive cost \(c\) with exact
purchases, a deterministic algorithm has ratio
\((1+\sqrt2)(1+\sqrt2-1/\rho)\) on arbitrary lengths. A separate
algorithm has ratio \(2\rho\) without an announced range.
On a finite undirected graph with nonnegative rational input data,
rooted Steiner service admits a polynomial-time deterministic algorithm
of ratio \(3+2\sqrt2\).
\end{theorem}

The single-edge TCP subinstance gives the equal-length lower bound.
For the \(2\rho\) policy, give each job at arrival a permanent mark
equal to the largest processing length seen so far. Use the largest
mark in the batch in place of the common length in the preceding timer.
Each launch again leaves enough time after its predecessor. The extra
cost in the offline comparison is controlled by the spread
(Appendix~\ref{app:price}).

\paragraph{Pricing a long occupation.}
For arbitrary lengths, run the instantaneous threshold virtually with
\(h(B)=c(B)+\lambda P(B)\) and merge its purchases at the next free
instant. A batch can be blocked by its predecessor and must finish its
own jobs. Summed over physical launches, each duration is thus charged
at most twice. If \(b\) indexes the virtual purchases with costs \(c_b\),
durations \(P_b\) and maximum waits \(W_b\), this gives
\[
 \ALG\le\sum_b(c_b+W_b+2P_b).
\]
The virtual timer fires precisely at \(W_b=c_b+\lambda P_b\). Since
\(\sum_b W_b\le\OPT_h\) by its factor-two guarantee, the last display
is at most \((1+2/\lambda)\OPT_h\) for \(1\le\lambda\le2\).
In an offline batch of maximum flow \(M\), the longest job gives
\(P\le M\) and an oldest job gives \(W+P/\rho\le M\).
Its virtual cost at the same launch is therefore at most
\(c+(1+\lambda-1/\rho)M\). Taking \(\lambda=\sqrt2\) balances the
two factors at unrestricted spread.

\paragraph{A direct timer on the line.}
Let \(x_j\) be job \(j\)'s distance from the root. Start all pending
jobs when the processor is free and the oldest wait reaches or exceeds
\(\max_j(x_j+p_j)\) over those jobs. If distance and processing
length are ordered together, a single job attains both maxima.
Its backward interval then pays for the batch exactly as in the TCP
timer analysis, giving competitive ratio five for arbitrary lengths.
When the two maxima belong to different jobs, the guarantee depends
on how much their sum exceeds \(\max_j(x_j+p_j)\).
Theorem~\ref{thm:lad-paired-timer} quantifies this dependence.

\paragraph{Computing the price on a graph.}
Let \(L(B)\) be the rooted Steiner cut LP value for the terminals of
\(B\). Adding terminals adds constraints, and summing two feasible
vectors covers the cuts required by their union. Thus \(2L\) is
monotone and subadditive. The primal-dual algorithm
of~\cite{gw} buys an actual tree for at most \(2L\), and \(L\le c\)
because a minimum tree is LP feasible. Use \(2L+\lambda P\) for the
virtual threshold and pool the supplied trees. This changes the
offline comparison to \(2c+(1+\lambda-1/\rho)M\). With
\(\lambda=\sqrt2\) the unrestricted bound is again \(3+2\sqrt2\);
with \(\lambda=2\) the ratio is \(6-2/\rho\). Appendix~\ref{app:price}
gives the LP and implementation details.

\section{Online service with delay}\label{sec:osd-main}
In OSD a server starts at a specified point of a known metric. A service
uses a walk from its current point through the locations of the selected
jobs and pays the walk length~\cite{osd,osdmax}. In our job model the
walk takes place at launch. Each selected job then runs for its own
processing time, and the next walk must wait until the processor is
free. The server's new position is the endpoint of the walk. Service
is elective, so passing a location leaves the algorithm free to defer
other pending jobs there.

Consider two buffered walks, one from \(x\) to \(y\) and one from \(y\)
to \(z\). The set of visited points specifies neither the starting
point required by the first walk nor the endpoint needed by the next
one. Returning to \(x\) after every purchase could repeatedly add the
cost of a long return trip. Instead the physical server can execute
\(x\) to \(y\) and then \(y\) to \(z\) in a single launch, admit all the
assigned jobs, and finish at \(z\). This construction is useful for
both objectives. Their proofs differ in how they pay for waiting
between consecutive walks.

\subsection{Total flow time}\label{sub:osd-total}
Run the instantaneous OSD algorithm on the requests. Record the walk
leading up to each nonempty service, including any movements since the
previous one. While the physical processor is busy the virtual server
continues from its own position. At the next opportunity execute all the
recorded walks in chronological order. This rule gives an invariant:
the physical position after each launch is the endpoint of the last
virtual walk executed. The first new walk in the buffer therefore
starts exactly where the physical server is waiting.

\begin{theorem}[Total flow for elective OSD]\label{thm:osd-front}
An instantaneous algorithm of ratio \(\alpha\) for movement plus total
waiting yields ratio \(\max\{\alpha,1+\rho\}\) for movement plus total
flow on the shared drop-line processor.
\end{theorem}

\paragraph{Proof sketch.}
The endpoint invariant allows each virtual leg to be used once.
For example, a movement with no assigned job can be retained with the
next service, so moving through an empty region creates no extra job
batch. The total physical movement is at most virtual movement \(N_V\).
A virtually served job again waits for at most the end of the current
physical occupation. Erasing processing from an offline schedule still
gives a valid comparison: its successive walks have precisely their
original starting positions. Thus with the notation of
Section~\ref{sub:rooted-total},
\[
   \ALG_{\rm sum}\le N_V+G_V+(1+\rho)P_\Sigma,\qquad
   \OPT_0+P_\Sigma\le\OPT .
\]
The first inequality uses the online walk order and the second uses
the offline walk order. The two servers need never be at the same point.
The result follows by applying the request guarantee to \(N_V+G_V\).
This is where the model's distinction between movement cost and elapsed
time enters. Traversing the recorded walks consumes no processor time
beyond the occupation of the jobs they launch.

\subsection{The sum of batch maximum flow times}\label{sub:osd-max}
A mobile request algorithm prices a walk from its current position.
Adding \(2P(B)\) to a static set cost, as on a rooted tree, does not
give the price of the next walk from its new position. We keep the
request algorithm's movement rule and its original objective. The
proof must then pay for the durations of all physical walks' batches
although the virtual algorithm has paid for none of them.

\begin{theorem}[Batch maxima for elective OSD]\label{thm:osd-max}
An instantaneous algorithm of ratio \(\alpha\) for movement plus the
sum of per-service maximum waiting times yields a
\((3\alpha+6)\)-competitive algorithm for movement plus the sum of
batch maximum flow times.
\end{theorem}

\paragraph{Which quantity needs a new bound?}
Let \(W_V\) be the sum of virtual maximum waits and \(P_i\) the duration
of the \(i\)th physical batch. The calculation from rooted aggregation
still bounds a batch maximum by virtual waiting plus \(P_{i-1}+P_i\).
The endpoint invariant pays for movement. Consequently
\[
                  \ALG_{\rm batch}\le N_V+W_V+2\sum_i P_i.
\]
For an optimal physical schedule write \(F^*=\sum_H M_H^*\) for its
sum of maximum flows. Picking a longest job in each online batch gives
a tempting comparison to \(F^*\): the optimum must pay at least that
job's length. The difficulty is multiplicity. Several of the chosen
jobs can lie in the same offline batch, whose maximum is paid only once.

\begin{figure}[htbp]
\centering
\begin{tikzpicture}[x=.75cm,y=.85cm,font=\small]
  \draw[guide!45] (5,-.22)--(5,2.88);
  \draw[->,guide] (-4.4,0)--(10.7,0) node[right] {\scriptsize time};
  \foreach \t in {-4,0,1,4,5,8,10}{
    \draw[guide] (\t,0)--(\t,-.12);
    \node[below] at (\t,-.12) {\scriptsize \(\t\)}; }
  \node[left] at (-4.45,2.5) {Comparison};
  \filldraw[fill=inkblue!12,draw=inkblue] (-4,2.22) rectangle (5,2.78);
  \node at (.5,2.5) {\scriptsize \(I_H=[-4,5]\)};
  \node[left] at (-4.45,1.65) {Virtual service};
  \foreach \t/\job in {0/A,1/B,5/C}{
     \fill[inkblue] (\t,1.5) circle (1.8pt);
     \node[above] at (\t,1.53) {\scriptsize \(\job\)}; }
  \node[left] at (-4.45,.67) {Occupation};
  \filldraw[fill=inkblue!18,draw=inkblue] (0,.43) rectangle (4,.91);
  \filldraw[fill=ochre!14,draw=ochre] (4,.43) rectangle (8,.91);
  \filldraw[fill=ochre!14,draw=ochre] (8,.43) rectangle (10,.91);
  \foreach \t/\job in {2/A,6/B,9/C}{
     \node at (\t,.67) {\scriptsize \(\job\)}; }
\end{tikzpicture}
\caption{Two occupations can escape one component. At a single metric
point, jobs \(A,B,C\) have release times \(0,1,5\) and lengths
\(4,4,2\). The virtual server serves each immediately. A feasible
comparison launches all three at time \(5\), paying maximum flow \(9\).
Its interval \(I_H\) contains all three virtual service times. Only
the first physical occupation fits inside it. The second crosses its
right endpoint, and the third starts beyond it.}
\label{fig:osd-component}
\end{figure}
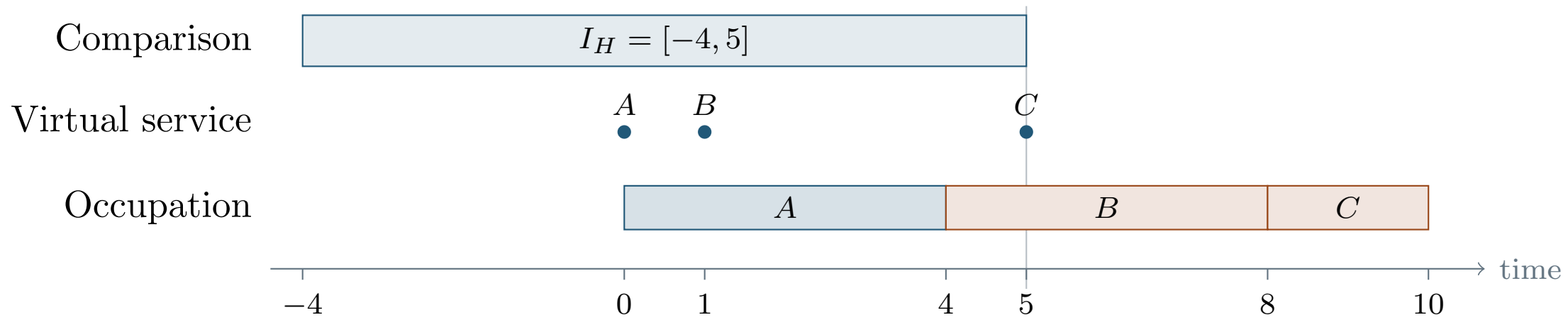

\paragraph{Proof sketch.}
For a comparison batch \(H\) launched at \(t_H\), draw the interval
\(I_H=[\min_{j\in H}(r_j-p_j),t_H]\). Its length equals \(M_H^*\).
For a virtual event \(b\) at time \(a_b\), draw its waiting interval
\(J_b=[\min_{j\in b}r_j,a_b]\). The union of the two families has
length at most \(F^*+W_V\). Now take a longest job of physical batch
\(i\). Its comparison interval reaches from at least one processing
length before its release up to that release, and its virtual interval
extends from the release to virtual service. They meet at the release,
so they determine a single connected component to which we assign \(i\).
Figure~\ref{fig:osd-component} shows a batch assigned to a component
even though its physical launch takes place beyond the right endpoint.

Why can that component pay for several batches? Let \([L,R]\) be a
component and \(q\) its last assigned physical batch index. If \(q\ge2\),
the representative of \(q\) received virtual service no earlier than
the launch of batch \(q-1\), since that launch cleared the buffer.
Its virtual service lies in \([L,R]\), so batch \(q-1\) started by \(R\).
Every assigned batch with index at most \(q-2\) started after or at \(L\)
and finished by the start of \(q-1\). Their occupied intervals thus
fit disjointly inside \([L,R]\), even if the assigned indices have gaps.
Only indices \(q-1\) and \(q\) can escape this accounting. Each such
duration is bounded by its representative's offline batch maximum.
That offline interval belongs entirely to this component, so it can
receive at most two exceptional charges. Summing gives
\[
            \sum_iP_i\le (F^*+W_V)+2F^*=W_V+3F^*.
\]
Substituting in the preceding cost bound gives
\(\ALG_{\rm batch}\le 3(N_V+W_V)+6F^*\). The offline physical walks
at their original starts give a feasible instantaneous schedule whose
waiting is at most \(F^*\). Thus \(N_V+W_V\le\alpha\OPT\), and
\(F^*\le\OPT\). Appendix~\ref{app:osd} gives the full proof.

On finite lines, weighted trees, and finite metrics, respectively,
the instantaneous guarantees \(10,12,20\) of~\cite{osdmax} yield
deterministic job guarantees \(36,42,66\). The metric enters through
the request algorithm. The bound on occupied time uses only the
chronological order of the walks and the two families of time intervals.

\FloatBarrier
\section{Facility location with delay}\label{sec:facility-main}
Suppose a region has two service sites a distance \(L\) apart, each
with opening fee one, and one client at each site. Serving each client
locally costs two in total. If the processor is required to serve both
clients in one opening, it pays \(1+L\). For arbitrarily large \(L\),
no constant factor relates this merged purchase to the two local ones.
The obstruction persists even if both clients are already waiting
and all distances form a metric.

In the ephemeral FLD model an opening serves the clients assigned at
that instant~\cite{azar-touitou}. Our sites share one dispatch resource. Serving clients at one site
occupies that resource even when every other site is idle. There are \(k\) candidate
facilities, with opening charge \(o_f\) and connection charge
\(\ell_{fj}\) from site \(f\) to job \(j\). A physical batch serves
one site group \(B\) at cost \(g_f(B)=o_f+\sum_{j\in B}\ell_{fj}\).
The dispatch resource becomes free when the longest job at that opening finishes.
A subsequent opening, even at the same site, pays again. Both objectives
use the same physical rule for executing assignments. We present it
once and then analyze what each delay objective charges for a round.

\paragraph{A round of openings.}
When the processor becomes free, collect all outstanding virtual
assignments and freeze their membership. Merge the jobs assigned to
the same site. Process the resulting nonempty site batches in a fixed
order, one launch per site. Assignments made during this round go
into the next round, even if their site has yet to open in the
current round. When the buffer is empty the processor waits for the
next virtual event. Freezing the round bounds its length and gives
every buffered job a definite round in which it will start.

For example, suppose the frozen site list is \(f,g\) and its longest
lengths there are \(p,q\). A new virtual assignment made just after
\(f\) starts waits almost \(p+q\) for the next round. If it is assigned
to \(g\) and the next round opens \(f\) first, it waits through that
opening as well. No client changes its assigned facility in this
construction. The only spatial effect of pooling is to save repeated
opening fees at the same site.

\begin{theorem}[Ephemeral facilities]\label{thm:fld-front}
For total flow, a deterministic \(\alpha\)-competitive instantaneous
algorithm for the single-facility-per-opening request problem gives
competitive ratio \(\max\{\alpha,1+(2k-1)\rho\}\).
For the sum of batch maxima there is a deterministic algorithm of
ratio \(k(6-2/\rho)\) with an exact static assignment oracle.
A second choice of timer gives \(k(3+2\sqrt2)\) for every spread.
Every deterministic algorithm has ratio at least \((k+1)/2\)
over instances of unrestricted spread.
The upper bounds use an assignment oracle with running time polynomial
for each fixed \(k\). The lower bound holds even on a uniform metric.
\end{theorem}

\paragraph{Total flow.}
Use the instantaneous total-wait algorithm to choose the assignments.
A round of at most \(k\) site batches lasts at most \(kp_{\max}\).
Once its current round ends, a buffered job encounters at most \(k-1\)
batches ahead of its own site in the next round. Thus the extra wait
from virtual service to physical launch is at most \((2k-1)p_{\max}\).
All physical connection charges occur in the virtual service, and
each site opening is paid for by an opening at the same site. The
additive argument now gives
\[
  \ALG_{\rm sum}\le \alpha\OPT_0+
       \bigl(1+(2k-1)\rho\bigr)P_\Sigma,\qquad
       \OPT_0+P_\Sigma\le\OPT .
\]
When \(k=1\) no other site can precede the job in the next round, and
this coefficient reduces to the one-batch blocking term for TCP.

\paragraph{A virtual assignment for batch maxima.}
One oldest wait can be charged at several physical openings. For
example, if two sites each receive jobs released at time zero, launching
the two groups separately after time \(t\) incurs at least \(t\) in
each group's maximum. A virtual event placing them both at time \(t\)
would have only one oldest wait. We therefore give a virtual event with
job set \(B\) and oldest wait \(W\) the price \(h(B)+kW\), where
\[
 h(B)=\min_{B=\dot\bigcup_f B_f}
        \sum_{f:B_f\ne\varnothing}
             \left(g_f(B_f)+2kP(B_f)\right).
\]
This virtual minimum allows several sites even though each physical
opening uses one. Restricting an assignment to fewer clients can only
reduce its cost. Combining two assignments at each site pays for the
opening once and takes the larger of two longest lengths. Thus \(h\)
is monotone and subadditive. The instantaneous threshold applied to
\(h/k\) is a factor-two algorithm for the displayed price after
multiplying its objective by \(k\). Keep the minimizing site assignments
it produces, and execute them with the same rounds as above.

\paragraph{Proof sketch for batch maxima.}
Let \(D_i\) denote the sum of the site durations in round \(i\), and
let \(W_i\) be the largest oldest wait among its virtual events.
Every physical star in that round has maximum flow at most
\(W_i+D_{i-1}+D_i\), with \(D_0=0\). The preceding round can block
the event for \(D_{i-1}\). Once this round starts, the time until even
the last site finishes is at most \(D_i\). There are at most \(k\)
physical maxima to charge. Summing over rounds charges any duration
\(D_i\) at most \(2k\) times, through its own round and its successor.
This explains both coefficients in \(h(B)+kW\) and shows that total
virtual cost pays for the physical execution.

For comparison take a single offline opening of cost \(g\), longest
length \(P\), oldest wait \(W\), and maximum flow \(M\). Its site is
an admissible assignment in the minimization for \(h\). As on a rooted
tree, \(W\le M-P/\rho\) and \(P\le M\). Its virtual price satisfies
\[
  h(B)+kW\le g+2kP+kW
     \le g+k(3-1/\rho)M
     \le k(3-1/\rho)(g+M).
\]
Erasing occupation makes all these virtual events feasible. The
instantaneous factor of two completes the comparison.

For fixed \(k\), the oracle can guess a maximum length for each site
from \(\{0\}\cup\{p_j:j\in B\}\), with zero denoting a closed site.
For each guess assign jobs to their cheapest eligible site and evaluate
the actual openings and site maxima. The vector of an optimal
assignment occurs among the guesses. Appendix~\ref{app:facility}
proves exactness and gives the running time. The factor \(k\) in the
competitive bound comes from how many site batches a round may
require. A long job followed by clients at all \(k\) sites shows that a linear
dependence is necessary at unrestricted spread. Give each site opening
cost one, place the sites far apart, and release one job of length \(P\).
Just after the algorithm commits to it, send a unit job to each site.
Serving locally forces \(k\) distinct batches to wait through nearly
all of the same occupation. Offline either finishes the long job before
this burst or serves the short jobs first, paying at most
\(2P+O(k^2)\). An expensive nonlocal assignment cannot evade the
lower bound. As \(P\) grows this gives a deterministic lower bound of
\((k+1)/2\), already with two distinct lengths
(Appendix~\ref{app:facility}).

\section{Concluding remarks}\label{sec:conclusion}
For total flow, each job contributes its own processing length to the
offline cost. For batch maxima, a whole batch may contribute only one
such length. What happens during an occupied interval then matters:
a rooted service can pool postponed purchases, a walk must preserve
their order, and several facilities can each incur the same wait.

\paragraph{How much does occupation cost at one root?}
The deterministic TCP batch-maximum ratio lies between \(2+2\sqrt2\)
and five at unrestricted lengths. The upper bound is tight for the
pending timer. Could a rule use release times and lengths jointly
to choose its next launch, or serve only part of the pending set?
The train construction tests the price of waiting for another long
job against the residual occupation seen by a short one.

\paragraph{Which losses come from the network?}
On a line, the direct timer achieves five when radii and lengths are
ordered together. Does five still suffice when the farthest and longest
jobs differ, or on a JRP tree? For OSD, can movement and occupation
be charged along the same walk to improve \(3\alpha+6\)? For facilities,
how does the tight order \(\Theta(k)\) change at bounded spread?

\paragraph{Offline optimization.}
TCP admits an exact algorithm for both objectives with a fixed number
of distinct lengths. With arbitrarily many lengths, is either objective
polynomial-time solvable? For batch maxima, can a different way of
scheduling overlapping components improve the factor two?

\label{main-end}\par
\noindent\begin{minipage}{\textwidth}
\section*{Declaration of AI Use}
OpenAI's ChatGPT and Codex were used in developing this work, including
literature exploration, formulation and analysis of algorithms,
construction and checking of mathematical arguments, manuscript drafting
and revision, and preparation of LaTeX code, figures, and pseudocode.
AI-generated suggestions informed parts of the technical development
and exposition. The authors directed this process and take full
responsibility for the final manuscript, including the correctness of
its claims and proofs and the accuracy of its references.\par
\end{minipage}

\clearpage
\appendix
\section{Offline schedules}
\label{sec:offline}

When all releases are known, it is possible to group jobs using the time
they will occupy before paying any online penalty for guessing the next
arrival. We first solve a relaxation of the batch-maximum objective. Its
value serves a second purpose in Section~\ref{sec:max-upper}, where the
online algorithm must compete without knowing these groups.

\subsection{An interval formula for batch maxima}
For each job draw the interval \(I_j=[r_j-K-p_j,r_j]\), and let
\(U=\bigcup_j I_j\). The left endpoint may precede time zero. It
records a cost on the time axis. It places no requirement on when a job
can be processed. In the relaxation, every batch still pays its fee and
the maximum of its jobs' individual flow times, but one batch may start
while another is processing.

\begin{lemma}\label{lem:backward-interval-lower}
The minimum relaxed batch-maximum cost is \(|U|\), the length of the
union. It is attained by launching one batch at the right endpoint of
each connected component of \(U\).
\end{lemma}
\begin{proof}
A batch \(B\) starting at \(s\) covers its jobs' intervals by the
single interval \([\min_{j\in B}(r_j-K-p_j),s]\). Its length is exactly
\(K+\max_{j\in B}(s-r_j+p_j)\). Summing over batches gives the lower
bound. Conversely, for each component \([L,R]\) take all jobs whose
intervals it contains. Their latest release is \(R\), their smallest
left endpoint is \(L\), and launching them together at \(R\) costs
\(R-L\). The empty instance has zero cost.
\end{proof}

\begin{theorem}\label{thm:offline-max}
For arbitrary processing times there is an \(O(n\log n)\)-time offline
\(2\)-approximation for \eqref{eq:batch}. For equal processing times the
component schedule itself is an exact optimum in the same running time.
\end{theorem}
\begin{proof}
Sort the interval endpoints to find the components. Take their relaxed
batches in order of right endpoint. If the processor is free, launch
the next batch at its endpoint. Otherwise buffer every batch that becomes
due until the processor finishes, then launch their union. Include an
endpoint event at the finishing time in that union.

Consider a physical batch formed from components in a set \(G_i\).
Let their lengths be \(H_g\), and let \(P_i\) be the physical batch's
processing time, with \(P_0=0\). Each job starts at most \(P_{i-1}\)
later than in the relaxation. Its fee plus flow in the relaxed schedule
was at most its component length. Thus the physical batch costs at most
\(\max_{g\in G_i} H_g+P_{i-1}\). Also \(P_i\le\max_{g\in G_i}H_g\),
since each interval has length \(K+p_j\). Every component belongs to
one group, and hence
\[
 \operatorname{cost}_{\rm batch}
 \le \sum_i\max_{g\in G_i}H_g+\sum_iP_{i-1}
 \le 2|U|\le 2\OPT .
\]
Grouping and replay take linear time after sorting.

For a common length \(p\), set \(d=K+p\). If \(a\) is the first release
of the next component and \(R\) the right endpoint of the current one,
then \(a-d>R\). The next relaxed launch is at least \(a>R+p\).
The component schedule therefore respects occupation and attains its
lower bound. In particular, for sorted releases of a nonempty instance,
\[
 \OPT=d+\sum_{j=2}^{n}\min\{r_j-r_{j-1},d\}.
\]
Each consecutive interval adds exactly its truncated release gap to the
union. Ties add zero.
\end{proof}

Occupation can make the relaxation strict. With \(K=1\) and jobs
\((r,p)=(0,4),(3,1)\), its two components have total length seven.
The relaxed launches at zero and three conflict. Combining the jobs at
three costs eight, as does serving the long job at zero and the short
one at four. These are the cheapest schedules in their respective
orders, and serving the short job first delays the long one still more.
Thus the physical optimum is eight.

\subsection{An exact algorithm for equal-length total flow}
\begin{proposition}\label{prop:offline-sum}
With common processing time \(p\), the offline total-flow objective
\eqref{eq:sum} can be minimized using \(O(n^4)\) arithmetic operations.
\end{proposition}
\begin{proof}
Sort jobs by release and fix an optimal sequence of launch times.
Moving an available job to the earliest of those launches at which it
is eligible can only reduce its flow. Because all lengths equal \(p\),
adding it never extends a batch. Empty launches may be removed. Thus an
optimum partitions the release order into consecutive blocks.
For a fixed partition, shift each launch as far left as its last release
and the previous finish permit. This preserves the partition and cannot
increase cost. Each finish then belongs to the grid
\(\mathcal T=\{r_j+kp:1\le j\le n,\ 1\le k\le n\}\), by induction
along the sequence of blocks.

Store the least cost of processing the first \(i\) jobs and finishing
at time \(t\), initialized only at \((0,0)\) with cost zero.
For each finite entry and each \(h>i\), start jobs \(i+1,\ldots,h\)
at \(s=\max\{t,r_h\}\). Relax the entry \((h,s+p)\) with the old cost
plus
\[
 K+(h-i)(s+p)-\sum_{j=i+1}^{h}r_j.
\]
All reachable finishes stay on the grid: the number of successive
increments of \(p\) is at most the number of jobs used. Every transition
is feasible, and the left-shifted optimum follows such transitions.
The answer is the minimum entry at \(i=n\). Prefix sums give constant
work per transition, with \(O(n^3)\) states and at most \(n\) successors
apiece. For an empty instance return zero.
\end{proof}

\subsection{A fixed number of distinct lengths}\label{app:offline-types}
\paragraph{Offline TCP with \(d\) distinct processing lengths.}
For either objective there is an exact \(n^{O(d)}\)-time algorithm.
Fix a total order by release time and identity in each length class.
There exists an optimum that processes each class in this order.
For batch maxima, if a later-ranked job runs before an earlier-ranked
job of the same length, postpone the former into the latter's batch.
It can start there, does not extend the batch, and its flow is at
most the latter job's flow. The old batch can only decrease its cost.
Keep the original launch slots, including emptied slots, until the
repairs finish: the sum of occupied slot indices over jobs strictly
increases and is bounded. For total flow, sort each class's jobs
over its occupied slots. Each swap of an adjacent inversion is
feasible at both launches, preserves both durations and the sum of
the two flows, and leaves the number of fees fixed.

For a fixed ordered partition, the best starts are the earliest
feasible ones. Every noninitial finish is consequently of the form
\(r_j+\sum_h m_hp_h\), where \(p_h\) ranges over the \(d\) distinct
lengths and the nonnegative integer coefficients sum to at most \(n\).
Indeed each start either equals the preceding finish or is set by a
release. Store minimum partial cost indexed by the \(d\) served prefix
sizes and the previous finish (initially zero). Choose a larger
componentwise prefix vector, with equality permitted in individual
coordinates. The difference specifies nonempty batch \(B\); launch at
\(s=\max\{t,\max_{j\in B}r_j\}\), and update the finish to \(s+P(B)\).
Add \(K+s+\max_{j\in B}(p_j-r_j)\) for batch maxima or
\(K+|B|s+\sum_{j\in B}(p_j-r_j)\) for total flow. Take the cheapest
final state. There are at most \((n+1)^d(1+n(n+1)^d)\) states and
\((n+1)^d\) transitions per state. This is polynomial for each fixed
\(d\) and permits unbounded numerical spread. The general-length
complexity for unbounded \(d\) is not settled by the recurrence.

\paragraph{Tight example for the existing offline replay.}
Take \(K=1\) and jobs \((r,p)=(0,P),(3,1)\) with \(P>3\).
The intervals are disjoint. The replay starts at \(0,P\) and pays
\((P+1)+(P-1)=2P\). Launching both at \(3\) costs \(P+4\), so the
ratio is at least \(2P/(P+4)\to2\). This proves tightness for that
replay algorithm. It leaves room for better offline algorithms.

\section{Total flow time}
\label{sec:total}
For total flow, a job's own processing contributes exactly \(p_j\)
regardless of when it starts. This fixed contribution and the cost of an
instantaneous acknowledgment instance give a joint lower bound.
Algorithm~\ref{alg:virtual} uses one clock for the virtual process and
stores virtually acknowledged jobs until the root is free. An arrival
changes the rate at which virtual waiting accumulates.

\begin{algorithm}[htbp]
\caption{Virtual acknowledgments for total flow time}\label{alg:virtual}
\begin{algorithmic}[1]
\State $V\gets\varnothing$ \Comment{Jobs awaiting a virtual acknowledgment}
\State $D\gets\varnothing$, $E\gets0$ \Comment{Jobs cleared virtually, and next free time}
\State Wake on arrivals, the current virtual deadline, and physical batch ends.
\AtTime
\State Add all jobs arriving at $t$ to $V$.
\If{$V\ne\varnothing$ and $\sum_{j\in V}(t-r_j)\ge K$}
  \State $D\gets D\cup V$ and $V\gets\varnothing$.
\EndIf
\If{$t\ge E$ and $D\ne\varnothing$}
  \State Launch $D$ at $t$, set $E\gets t+P(D)$, then empty $D$.
\EndIf
\If{$V\ne\varnothing$}
  \State Set the virtual deadline to $(K+\sum_{j\in V}r_j)/|V|$.
\Else
  \State Cancel the virtual deadline.
\EndIf
\end{algorithmic}
\end{algorithm}

\begin{theorem}\label{thm:total}
For each \(K>0\) and each \(\rho\ge1\), the optimal deterministic competitive
ratio for \eqref{eq:sum} on inputs of spread \(\rho\) is \(1+\rho\).
One algorithm attains the bound simultaneously for all spreads.
\end{theorem}

\begin{proof}
Let \(P_\Sigma=\sum_jp_j\), and let \(Q\) be the minimum fee plus total
waiting time when acknowledgments are instantaneous and the release times
are unchanged. A physical schedule that starts job \(j\) at \(s_j\)
pays \(P_\Sigma\) plus its fees and \(\sum_j(s_j-r_j)\). Its batch
starts are feasible instantaneous acknowledgment times, so
\begin{equation}\label{eq:sum-lb}
  P_\Sigma+Q\le\OPT.
\end{equation}

Simulate an instantaneous acknowledgment policy on the original arrivals.
Send a virtual acknowledgment for all virtually pending jobs as soon as
their total waiting time reaches \(K\). For completeness, each virtual
phase pays exactly \(2K\). In the interval ending at its acknowledgment
and excluding the preceding acknowledgment, an instantaneous comparator
either sends an acknowledgment (paying \(K\)) or leaves all jobs of the
phase waiting until the phase ends (incurring at least \(K\)).
For the first phase include its first release. Arrivals at a virtual
acknowledgment time belong to the phase ending there. The charged
acknowledgments are distinct, and waiting charges concern disjoint sets
of jobs. The virtual cost \(V\) is therefore at most \(2Q\).

Continue the simulation while the physical processor is busy. Whenever
the processor is free and there are buffered virtual acknowledgments,
launch their union as one physical batch. Include an acknowledgment
issued exactly at a batch end before forming the next batch.
The physical schedule pays no more fees than the simulation. If \(a_j\)
is the virtual acknowledgment time, then \(0\le s_j-a_j\le p_{\max}\):
the next time the processor is free, all buffered jobs start together.
For \(n\) jobs, \eqref{eq:sum-lb} now gives
\[
 \ALG\le V+P_\Sigma+n p_{\max}
      \le 2Q+(1+\rho)P_\Sigma
      \le (1+\rho)\OPT.
\]

For the converse, release two jobs of lengths \(1\) and \(\rho\) at zero,
and designate the latter as the long job (also when \(\rho=1\)).
Every feasible deterministic algorithm eventually starts it at some
finite \(T\) on this prefix. At \(R=T+\varepsilon\), with
\(0<\varepsilon<\rho\), release \(m\) additional unit jobs. Each has
flow at least \(\rho+1-\varepsilon\). Offline can start all jobs at \(R\)
in one batch of cost \(K+m+1+\rho+2R\). For fixed \(\varepsilon\), let
\(m\) tend to infinity, then let \(\varepsilon\) decrease to zero.
The initial two jobs fix the exact spread on both the prefix and its
extension, even if the algorithm knows the range in advance.
\end{proof}
\FloatBarrier

\section{A deterministic upper bound for the maximum-flow objective}
\label{sec:max-upper}

For objective~\eqref{eq:batch}, the age of the oldest pending job gives
a natural clock. The length of the longest pending job determines both
the clock's threshold and the occupation time of the batch at launch.

\subsection{The Pending timer algorithm}

The \emph{Pending timer} algorithm starts all currently pending jobs
whenever it starts a batch.  Here pending jobs are jobs not yet assigned
to a batch, rather than jobs already running in the processor.  When
the pending set first becomes nonempty, let $b$ be its earliest release
time and set a timer for
\[
  b+K+\max\{p_j:j\text{ is pending}\}.
\]
Every subsequent arrival before the next start updates the maximum and
the timer.  In particular, an arrival may postpone this deadline.  The
algorithm starts the pending set as soon as the processor is free and
the current timer has expired.  An arrival exactly at a tentative
deadline is included before the timer is tested.  The algorithm uses no
processing-time ratio, future job count, or end-of-input notification.
Algorithm~\ref{alg:timer} gives the event order. It suffices to schedule
one wake-up at the later of the timer threshold and the next free time.

\begin{algorithm}[htbp]
\caption{Pending timer for the sum of batch maximum flow times}\label{alg:timer}
\begin{algorithmic}[1]
\State $A\gets\varnothing$, $E\gets0$ \Comment{Pending jobs and next free time}
\State Wake on arrivals and the current scheduled launch.
\AtTime
\State Add all jobs arriving at $t$ to $A$.
\If{$A\ne\varnothing$}
  \State $b\gets\min_{j\in A}r_j$, $P\gets\max_{j\in A}p_j$.
  \State $u\gets\max\{E,b+K+P\}$.
  \If{$t\ge u$}
    \State Launch $A$ at $t$, set $E\gets t+P$, empty $A$, and cancel the timer.
  \Else
    \State Replace the scheduled launch time by $u$.
  \EndIf
\EndIf
\end{algorithmic}
\end{algorithm}

Index its nonempty batches by $i=1,\ldots,q$.  Write $s_i$ for the
start time, $P_i$ for the longest processing time in the batch, and
$E_i=s_i+P_i$ for its occupied interval's end.  Also write
\[
  b_i=\min_{j\in B_i}r_j,\qquad
  d_i=K+P_i,\qquad w_i=s_i-b_i.
\]
Put $E_0=0$.  All intervals below are intervals of physical time;
their lengths are unaffected by the conventions at shared endpoints.

\begin{lemma}\label{lem:pending-timer-identities}
For every batch of Pending timer,
\[
  s_i=\max\{b_i+d_i,E_{i-1}\},\qquad
  w_i=d_i+\delta_i,
  \quad
  \delta_i=(E_{i-1}-b_i-d_i)_+.
\]
In particular $w_i\geq d_i$, $\delta_1=0$, and, for $i\geq2$,
\[
  \delta_i\leq(d_{i-1}-d_i)_+.
\]
If $C_i$ denotes the objective contribution of batch $i$, then
\[
  C_i\leq w_i+d_i=2d_i+\delta_i\leq2w_i.
\]
\end{lemma}

\begin{proof}
During the formation of a batch the earliest pending release stays
fixed, and the maximum pending processing time can only increase.
Consequently its tentative deadline can only move to the right.
If the batch starts at $E_{i-1}$, its final deadline is at most that
time.  Otherwise it starts when its final timer expires, after the
processor is already free.  These statements also hold if a deadline
coincides with an arrival: that arrival is processed first and, if it
extends the timer, the batch does not yet start.  Thus the final start
is exactly the displayed maximum.  Finitely many arrivals and finite
processing times ensure that the final pending set eventually starts.

Every previous start removed all jobs then pending, including all
arrivals at that timestamp.  Hence, for $i\geq2$, $b_i>s_{i-1}$.
It follows that
\[
  E_{i-1}-b_i\leq P_{i-1}=d_{i-1}-K\leq d_{i-1},
\]
which proves the overflow bound.  For the first batch,
$E_0-b_1\leq0$, so $\delta_1=0$.  Finally every job of batch $i$
satisfies $s_i-r_j+p_j\leq w_i+P_i$.  Therefore
$C_i\leq K+w_i+P_i=w_i+d_i$, as required.
\end{proof}

\subsection{An interval lower bound and component packing}

Recall the intervals \(I_j=[r_j-K-p_j,r_j]\) and their union \(U\) from
Section~\ref{sec:offline}. Its components are available only in hindsight.
We use them to account for the online timer's batches; the timer itself
uses only revealed jobs.

\begin{theorem}\label{thm:five}
Pending timer is a deterministic $5$-competitive algorithm for the
maximum-flow objective.  It does not require a bound on the
processing-time ratio.
\end{theorem}

\begin{proof}
For every algorithmic batch $i$, choose a longest job and denote its
release by $v_i$.  Its interval $[v_i-d_i,v_i]$ is contained in one
connected component of $U$.  Assign the batch to that component.
All jobs of the next batch arrive after the preceding start, and hence
\[
  v_{i+1}>s_i\geq v_i.
\]
Thus the batches assigned to any one component are consecutive in the
algorithm's batch order.  Components with no assigned batch may be
ignored throughout the upper-bound accounting.

Fix a used component $[L,R]$, write $H=R-L$, and relabel its consecutive
batches as $1,\ldots,k$ within this paragraph.  Every $d_i\leq H$,
and the first representative satisfies $v_1\geq L+d_1$.
For $2\leq i\leq k-1$, its waiting interval $[b_i,s_i]$ lies between
$v_1$ and $v_k$: indeed
\[
  b_i>s_{i-1}\geq v_1,
  \qquad s_i<b_{i+1}\leq v_{i+1}\leq v_k.
\]
These waiting intervals are disjoint, since $s_i<b_{i+1}$.  Therefore
\begin{equation}\label{eq:component-wait-packing}
  d_1+\sum_{i=2}^{k-1}w_i
    \leq d_1+v_k-v_1\leq H.
\end{equation}
For $k=2$ the empty sum has the usual interpretation and the inequality
still holds.

Suppose first that $k\geq2$.  Lemma~\ref{lem:pending-timer-identities}
and~\eqref{eq:component-wait-packing} bound the first and intermediate
batch costs by
\[
  C_1+\sum_{i=2}^{k-1}C_i
    \leq2d_1+\delta_1+2\sum_{i=2}^{k-1}w_i
    \leq2H+\delta_1.
\]
The batch immediately before the last one belongs to this same
component.  Since both of their $d$ values are at most $H$,
\[
  C_k\leq2d_k+(d_{k-1}-d_k)_+
       =d_k+\max\{d_k,d_{k-1}\}\leq2H.
\]
The component therefore contributes at most $4H+\delta_1$.
For a singleton component block the stronger estimate
$C_1\leq2H+\delta_1$ gives the same stated bound.  Here $\delta_1$
always denotes the overflow of the component's first batch, not a
resetting of the algorithm's actual state.

The first used component has zero initial overflow.  For every later
used component, the batch preceding its first batch belongs to the
previous used component.  Its $d$ value is at most that previous
component's length, so Lemma~\ref{lem:pending-timer-identities} charges
the new component's initial overflow to at most that length.  Each used
component receives at most one such charge, from its successor.
Consequently
\[
  \mathrm{ALG}
    \leq4\sum_{\text{used components}}H
       +\sum_{\text{component first batches}}\delta
    \leq5\sum_{\text{used components}}H
    \leq5|U|
    \leq5\,\mathrm{OPT},
\]
where the last inequality is
Lemma~\ref{lem:backward-interval-lower}.
\end{proof}

Figure~\ref{fig:packing} shows the disjoint intervals in
\eqref{eq:component-wait-packing} and a positive overflow into a
second component on the same instance.
\FloatBarrier

There are at most as many nonempty batches as jobs.  Maintaining the
first pending release, the maximum pending processing time, and the
processor's next completion suffices to implement the algorithm.
Rational input data give rational timer values with polynomial
encoding length.  This is an event-driven algorithm and does not
enumerate the numerical time horizon.

\subsection{Equal processing times: an optimal factor of two}

\begin{theorem}\label{thm:equal-two}
If all processing times equal a common $p>0$, Pending timer is
$2$-competitive.  No deterministic online algorithm without an
announced input length has competitive ratio strictly below $2$,
even when $K$ and $p$ are known.
\end{theorem}

\begin{proof}
Put $d=K+p$.  For every batch after the first,
$E_{i-1}-b_i<p<d$, so
Lemma~\ref{lem:pending-timer-identities} gives $s_i=b_i+d$; the same
identity holds for the first batch.  Its maximum flow is $d+p$, and
its contribution is exactly $2d$.

Take one job released at $b_i$ from each algorithmic batch.
The releases of successive representatives are more than $d$ apart.
An offline batch containing $k\ge1$ representatives incurs maximum
waiting at least $(k-1)d$ and pays the additional $K+p=d$.
Summing over offline batches gives $\OPT\ge qd$ for $q$ representatives.

For the lower bound, fix any deterministic algorithm that serves every
finite input.  Choose $0<\varepsilon<p$.  Release the first job at
$r_1=0$.  If $T_i$ is the time when the algorithm starts job $i$,
release job $i+1$ at
\[
  r_{i+1}=T_i+\varepsilon.
\]
Every job has length $p$.  The construction can be stopped at any
finite prefix; if some $T_i$ never occurs, that prefix already violates
feasibility.  No two chain jobs can share an algorithmic batch, since
the new job arrives strictly after the preceding start and before its
occupied interval ends.  In particular $T_i\geq T_{i-1}+p$.
On the prefix of $m\geq2$ jobs, its cost is at least
\begin{equation}\label{eq:equal-chain-alg}
  \sum_{i=1}^m(d+T_i-r_i)
     =md+T_m-(m-1)\varepsilon.
\end{equation}
One feasible offline schedule serves the entire prefix in a single
batch at $r_m$, for cost $d+T_{m-1}+\varepsilon$.
Another merges the first two jobs at $r_2$ and starts every subsequent
job at its own release.  This is feasible because
$r_3\geq r_2+p$ and $r_{i+1}-r_i\geq p$ for $i\geq2$.
Its cost is $(m-1)d+T_1+\varepsilon$. Summing the costs of the two
comparators gives
\[
  2\OPT_m\le md+T_m+T_1+2\varepsilon .
\]
Together with~\eqref{eq:equal-chain-alg} this implies
\[
 \liminf_{m\to\infty}\frac{\ALG_m}{\OPT_m}
 \ge 2(1-\varepsilon/d),
\]
since $md+T_m\ge md$ and $T_1$ is fixed on all prefixes.
Letting $\varepsilon$ decrease to zero proves the lower bound.
\end{proof}

\section{Lower bounds for the sum of batch maximum flow times}
\label{sec:max-lower}

Stopping the sequence after a long job exposes the cost of its prefix.
For the unrestricted lower bound we append one short job when the
comparator is already free. That input pays for both the prefix and
the online processor's remaining occupation. Throughout this section the
algorithm has no advance bound on future processing times, and $K>0$
is fixed. The lower bounds allow an input-independent additive constant.

\subsection{A long-job prefix and its final release}

If each new job is long enough, launching promptly can force the
algorithm to pay for many separate batches. For this argument to work,
a comparator must still be able to wait for the last release and serve
everyone at little cost. At a common launch time \(s\), a job's flow
is \(s+(p_j-r_j)\). Choosing equal values of \(p_j-r_j\) therefore makes
the price of the offline batch independent of the number of jobs it
contains. It also leaves a long occupation after the algorithm's last
launch, which the other continuation can exploit.

Fix $0<\epsilon<1$.  Release the first job with $r_1=0$ and $p_1=1$.
Whenever the algorithm starts job $i$ at time $T_i$, the adversary may stop,
or release the next job with
\begin{equation}
 \label{eq:max-long-train}
 r_{i+1}=T_i+\epsilon,
 \qquad p_{i+1}=r_{i+1}+1.
\end{equation}
Thus each algorithmic batch contains exactly one job.  Put
\[
 H=K+1,\qquad b_i=T_i+H,\qquad S_i=\sum_{j=1}^i b_j.
\]
The algorithm's cost on the first $i$ jobs is at least $S_i$, since
$p_j-r_j=1$ for every job.  The starts $T_i$ exist by finite-input
completion.  They are strictly increasing and tend to infinity, because
successive batches cannot overlap and all processing times are at least one.

At launch \(T_i\), \(i\ge2\), stopping gives a simple comparison:
offline serves all train jobs at \(r_i=T_{i-1}+\epsilon\) for cost
\(b_{i-1}+\epsilon\). Their common flow is \(r_i+1\), and the longest
job finishes at \(r_i+p_i\). To charge the online processor for
blocking one more job, choose that job's arrival using this finishing
time.

\begin{lemma}\label{lem:max-two-forks}
Append a final unit job at \(R_i=\max\{T_i+\epsilon,r_i+p_i\}\).
A feasible offline schedule costs \(U_i=b_{i-1}+\epsilon+H\), whereas
the online cost is at least
\[
 S_i+\min\{b_{i-1},b_i-b_{i-1}\}.
\]
\end{lemma}
\begin{proof}
Launch the train offline at \(r_i\), then launch the unit job at
\(R_i\), when the processor is free. Online has already launched
job \(i\). Its remaining occupation at \(R_i\) is
\[
 T_i+p_i-R_i=\min\{p_i-\epsilon,T_i-r_i\}\ge0 .
\]
The additional fee plus flow is therefore at least
\[
 H+\min\{p_i-\epsilon,T_i-r_i\}
 =\min\{b_{i-1}+1,b_i-b_{i-1}+H-\epsilon\}.
\]
Since \(H\ge1>\epsilon\), this proves the bound.
\end{proof}

\begin{lemma}\label{lem:max-fork-sequence}
Let \(b_i\) be an increasing unbounded sequence of positive numbers,
\(S_i=\sum_{j=1}^i b_j\), and \(q_i=b_i/b_{i-1}\) for \(i\ge2\).
Then
\[
 \limsup_i \left(\frac{S_i}{b_{i-1}}+\min\{1,q_i-1\}\right)
 \ge 2+2\sqrt2 .
\]
\end{lemma}
\begin{proof}
Write \(q=\limsup_i q_i\). If \(q=\infty\), the first term is
unbounded. Otherwise take a subsequence with \(q_i\to q\).
For every \(\eta>0\), all sufficiently late ratios are at most
\(q+\eta\). Looking back \(m\) terms at sufficiently late indices gives
\[
 \frac{S_i}{b_{i-1}}\ge q_i+\sum_{h=0}^{m-1}(q+\eta)^{-h}.
\]
Take the lower limit on the subsequence, then let \(\eta\downarrow0\)
and \(m\to\infty\). For \(q=1\) the bound diverges. For \(q>1\) we
obtain \(f(q)=q^2/(q-1)\), the prefix-cost expression also seen in
online bidding~\cite{bidding}. The extra waiting term converges to
\(\min\{1,q-1\}\) on this same subsequence.
For \(q\ge2\), \(f(q)+1\ge5\). For \(1<q\le2\), put \(x=q-1\).
Then
\[
 f(q)+q-1=2+2x+1/x\ge2+2\sqrt2 ,
\]
with equality at \(x=1/\sqrt2\).
\end{proof}

\begin{theorem}\label{thm:max-unrestricted-lower}
Every deterministic online algorithm in the stated model has competitive
ratio at least \(c_*=2+2\sqrt2=4.8284271\ldots\), for every fixed
\(K>0\), even allowing an input-independent additive constant.
\end{theorem}

\begin{proof}
Use the infinite train to define its consistent finite prefixes. At
any chosen index \(i\) the terminal unit job replaces the next long
job after the same revealed history. No end-of-input signal is needed:
each \(T_i\) is determined by the algorithm on its finite prefix and
\(R_i>T_i\). Lemma~\ref{lem:max-two-forks} bounds its ratio to the
displayed comparator below by
\[
 \frac{S_i/b_{i-1}+\min\{1,q_i-1\}}
      {1+(\epsilon+H)/b_{i-1}} .
\]
The denominator tends to one. By Lemma~\ref{lem:max-fork-sequence},
for every \(0\le c<c'<c_*\), arbitrarily late finite inputs have
algorithm cost at least \(c'U_i\). As \(\OPT\le U_i\) and
\(U_i\to\infty\), their excess \(\mathrm{ALG}-c\OPT\) is unbounded.
Figure~\ref{fig:forks} displays the prefix and combined bounds.
\end{proof}

\subsection{The pending timer cannot improve on five}

Recall the pending timer of Theorem~\ref{thm:five}: if $b$ is the first
pending release, $P$ is the largest processing time among pending jobs, and
$E$ is the preceding batch's finishing time, it starts all pending jobs when
the current time reaches $\max\{E,b+K+P\}$.  Arrivals update $P$ before a
start decision.

\begin{theorem}
\label{thm:max-timer-tight}
For every fixed $K>0$, the pending timer has competitive ratio at least
$5$, even allowing an input-independent additive constant.  Together with
Theorem~\ref{thm:five}, its competitive ratio is exactly $5$.
\end{theorem}

\begin{proof}
Use the long train~\eqref{eq:max-long-train}, and append a final unit job
just after the timer starts train job $i$, for $i\ge2$.  Its starts obey
\[
 T_1=H,\qquad T_i=2T_{i-1}+H+2\epsilon\quad(i\ge2).
\]
Indeed, job $i$ arrives alone at $T_{i-1}+\epsilon$, and its own pending
deadline is $2T_{i-1}+H+2\epsilon$.  The preceding batch finishes no later
than that deadline, since $p_{i-1}\le T_{i-1}+1$ and $H>1$.
Thus the claimed start is reached before the next
train arrival.  In terms of $b_i=T_i+H$,
\[
 b_1=2H,\qquad b_i=2b_{i-1}+2\epsilon,
 \qquad
 b_i=2^{i-1}(2H+2\epsilon)-2\epsilon.
\]
In particular $b_i/b_{i-1}\to2$ and $S_i/b_{i-1}\to4$.

The final unit job arrives at $T_i+\epsilon$.  Its own deadline is
$T_i+\epsilon+H$, whereas the root is occupied until $T_i+p_i$.
For $i\ge2$, the latter is strictly later, since
\[
 p_i-\epsilon-H=T_{i-1}-K>0.
\]
The timer starts that unit job at the root's finishing time.  Its fee plus
flow time is exactly $b_{i-1}+1$, so the total algorithm cost is
$S_i+b_{i-1}+1$.

Offline, batch the train at its last release $r_i$.  This costs
$b_{i-1}+\epsilon$ and finishes at $2T_{i-1}+2\epsilon+1$.
The final unit job arrives strictly later: the difference is
$K+\epsilon>0$.  Serve it immediately for cost $H$.  Consequently
\[
 \frac{\mathrm{ALG}}{\mathrm{OPT}}
 \ge
 \frac{S_i+b_{i-1}+1}{b_{i-1}+H+\epsilon}
 \longrightarrow 5.
\]
The comparator cost tends to infinity, so every ratio strictly below five
also fails with an additive competitive constant.
\end{proof}

\section{Fixed processing-time spread}
\label{sec:fixed-spread}

We next consider the per-batch maximum objective on inputs of spread
\(\rho=p_{\max}/p_{\min}\).  The upper bounds below do not require the
algorithm to know either extreme or their ratio.  The lower bound uses
only two processing times and applies to the class of inputs whose
\emph{actual} spread is exactly \(\rho\).

\begin{theorem}
\label{thm:fixed-spread-upper}
The pending-job timer of Theorem~\ref{thm:five} has strict competitive
ratio at most
\[
 \min\left\{5,\frac{2(K+p_{\max})}{K+p_{\min}}\right\}
 \le \min\{5,2\rho\}.
\]
Neither the algorithm nor its choice of parameters uses \(\rho\).
\end{theorem}

\begin{proof}
Recall the rule: while the processor is idle, let \(b\) be the release
time of the oldest pending job and let \(P\) be the largest pending
processing time.  Launch all pending jobs as soon as
\(t-b\ge K+P\).  If the processor is occupied, test this condition again
when it becomes free.  New arrivals are incorporated before a launch,
and may increase the threshold.

Set \(H_{\min}=K+p_{\min}\) and \(H_{\max}=K+p_{\max}\), solely for the
analysis.  For algorithmic batch \(i\), let \(b_i\) be its first release,
\(s_i\) its launch time, and \(w_i=s_i-b_i\).  Then
\[
 H_{\min}\le w_i\le H_{\max}.
\]
The lower bound is the launch condition.  For the upper bound, its
threshold is always at most \(H_{\max}\).  If a preceding batch blocks
the processor at \(b_i\), that batch started no later than \(b_i\) and
finishes by \(b_i+p_{\max}\).  Hence both the processor and the timer
permit a launch by \(b_i+H_{\max}\), regardless of intervening arrivals.
Consequently each algorithmic batch costs at most
\[
 K+w_i+\max_{j\in B_i}p_j\le 2H_{\max}.
\]

Choose one job released at \(b_i\) as its representative.  Since each
launch clears all pending jobs and same-time arrivals precede a launch,
the first releases of successive algorithmic batches satisfy
\(b_{i+1}>s_i\ge b_i+H_{\min}\).  Consider any optimum batch containing
\(q\ge1\) representatives.  It starts no earlier than their last release.
The earliest representative therefore has flow at least
\((q-1)H_{\min}+p_{\min}\).  Including the batch fee, this optimum batch
costs at least \(qH_{\min}\).  Assigning every representative to its
optimum batch proves \(\OPT\ge mH_{\min}\), where \(m\) is the number
of algorithmic batches.  Thus \(\ALG\le2mH_{\max}\le
2H_{\max}\OPT/H_{\min}\).  The same timer is also
\(5\)-competitive by Theorem~\ref{thm:five}; no selection between
different algorithms is needed.  Finally,
\((K+p_{\max})/(K+p_{\min})\le\rho\).
\end{proof}

\begin{theorem}
\label{thm:fixed-spread-lower}
For every fixed \(\rho>1\) and every fixed fee \(K>0\), no deterministic
online algorithm has strict competitive ratio smaller than
\[
 R_\rho=1+\frac{2}{\sqrt{3+1/\rho}}
\]
on inputs of actual spread exactly \(\rho\).  The final number of jobs
and the end of input are unannounced.  The same lower bound holds if a
fixed additive competitive constant is allowed.
\end{theorem}

\begin{proof}
Write \(y=1/\rho\in(0,1)\), choose a scale \(P>0\), and use only the
lengths \(P\) and \(yP\).  We first describe an episode starting with an
idle processor and no pending jobs, taking its initial time as zero.
Release one job of length \(P\), and let \(S\) be the time at which the
algorithm launches it.  Fix \(\eta>0\) with \(\eta<yP\).  At time
\(S+\eta\), release a second job, of length either \(P\) or \(yP\).
It cannot join the running batch.  All bounds below are valid even if
the algorithm subsequently waits longer before its second launch.

If the second length is \(P\), the algorithm pays at least
\[
 A_{\rm equal}=2K+S+3P-\eta.
\]
A feasible offline schedule waits until \(S+\eta\) and launches both
jobs together.  Its cost is
\[
 B_{\rm equal}=K+S+P+\eta.
\]
If instead the second length is \(yP\), the algorithm pays at least
\[
 A_{\rm short}=2K+S+(2+y)P-\eta.
\]
An offline comparator processes the first job at time zero and the
second at \(\max\{P,S+\eta\}\), for cost
\[
 B_{\rm short}=2K+(1+y)P+(P-S-\eta)_+.
\]
These are feasible comparators, so dividing by their costs gives valid
lower bounds on \(\ALG/\OPT\).

Put \(x=S/P\).  As \(P\) tends to infinity with \(K\) and \(\eta\)
fixed, for \(0\le x\le1\) the two ratios converge uniformly to
\[
 f(x)=\frac{3+x}{1+x},\qquad
 g(x)=\frac{2+y+x}{2+y-x}.
\]
Here \(f\) is decreasing and \(g\) is increasing.  Their intersection is
\[
 x_0=\sqrt{3+y}-1\in(0,1),\qquad
 f(x_0)=g(x_0)=R_\rho.
\]
The adversary chooses the equal-length continuation when \(x\le x_0\)
and the short continuation when \(x>x_0\).  For \(x\ge1\), the short
comparator instead gives a ratio at least
\[
 \frac{2K+(3+y)P-\eta}{2K+(1+y)P}
 \longrightarrow \frac{3+y}{1+y}>R_\rho.
\]
Thus, for every \(\delta>0\), one sufficiently large \(P\) makes every
episode have algorithmic cost at least \((R_\rho-\delta)\) times its
chosen comparator cost, uniformly over the algorithm's choice of \(S\).

To ensure the same actual spread on every prefix used in the argument,
begin with two jobs released simultaneously, of lengths \(P\) and \(yP\).
Once the algorithm finishes both, run arbitrarily many episodes, releasing
the first job of the next episode only after all jobs of the current
episode have finished.  A feasible algorithm eventually finishes every
finite prefix; no notification that the prefix is final is supplied.
Thus the construction works even if the algorithm knows the exact two
lengths in advance and is promised that both occur.

Offline processes the initial pair at time zero in one batch, paying
\(K+P\) and finishing by the time the algorithm finishes both jobs.
Moreover, each episode comparator
finishes no later than the algorithm finishes that episode: for the
equal continuation its finish is \(S+\eta+P\le S+2P\), and for the
short continuation it is
\(\max\{P,S+\eta\}+yP\le S+P+yP\).
Hence the comparators concatenate feasibly at the actual episode
release times.  Their total cost grows without bound with the number
of episodes, while the initial pair contributes only once.  Letting that
number tend to infinity, and then \(\delta\) tend to zero, proves the
claim, including against any fixed additive constant.  All input
lengths are \(P\) or \(yP\), and both occur, so the spread is exactly
\(\rho\).  This is a deterministic adversary argument; no claim about
an oblivious-adversary randomized lower bound follows from it.
\end{proof}

For example, the lower bound is \(2.0690449\ldots\) at spread \(2\),
and tends to \(1+2/\sqrt3=2.1547005\ldots\) as the spread grows.

\section{Transfers to network aggregation}
\label{app:aggregation}

The buffering argument for total flow separates the network decision from
the occupation of the root.  This appendix states the resulting transfer
and the assumptions needed to apply it beyond a fixed acknowledgment fee.

\paragraph{Network and processing contract.}
Each job has a release time $r_j$, a revealed processing length $p_j>0$,
and a network requirement.  In the instantaneous model, a service action
can serve a chosen set of released requests.  In the physical model, the
same action admits that set as one unbounded drop-line batch: all its jobs
start at $s$, job $j$ finishes at $s+p_j$, and the root admits no new batch
before $s+P(B)$, where $P(B)=\max_{j\in B}p_j$.  The network action has
zero elapsed duration and a nonnegative cost independent of its execution
time.  It is a launch action, not a visit that must wait for the jobs'
processing completions.  The network state, if any, is separate from the
occupied root.  No batch capacity or forced admission of unassigned jobs
is imposed.

We require one of the following two realization properties.
\begin{enumerate}
\item In the stateless case, the union of supplied actions serving sets
  $B_1,\ldots,B_k$ can serve their union at cost at most the sum of the
  supplied costs.  In particular, an exact nonnegative, normalized,
  subadditive service cost $c$ has this property when its minimizing
  objects can be realized.
\item In the stateful case, any consecutive actions of a supplied
  service trace can be concatenated instantaneously in their original
  order.  The concatenation starts and ends in the same states as that
  segment, serves its assigned requests, and costs no more than the
  segment.  No other action changes the network state during root
  occupation.  Zero-time walks with distance charged as a cost have this
  property; walks with positive travel duration need not.
\end{enumerate}
These are properties of actual supplied actions.  The transfer does not
replace a virtual action by an unevaluated minimum-cost network object.

\begin{theorem}
\label{thm:merge-add}
Suppose the corresponding instantaneous network problem with unweighted
total waiting has a strict deterministic $\alpha$-competitive online
algorithm that supplies a finite executable action trace and serves every
request on every finite input.  Under either realization property above, there is a
deterministic online algorithm for network cost plus total flow on the
drop-line processor with strict competitive ratio
\[
 \max\{\alpha,1+\rho\},
 \qquad \rho=p_{\max}/p_{\min}.
\]
It needs no advance knowledge of the processing-time range or the end of
input. If action unions or ordered concatenations can be constructed
in polynomial time, the additional work is polynomial in the supplied
trace and the input size.
\end{theorem}

\begin{proof}
Simulate the instantaneous algorithm on the original releases, independently
of the physical schedule.  Assign each request to the virtual action that
first serves it.  Execute an action immediately when the root is free.
While the root is occupied, buffer all virtual actions; at the next root
completion, merge or concatenate all buffered actions and admit their
assigned jobs together.  If there is no buffered action, wait for the next
one.  All virtual actions at a common timestamp are incorporated before
the physical launch at that timestamp.  All arrivals at that timestamp
are first revealed to the virtual algorithm.

Each physical group consists of consecutive virtual actions.  In the
stateful case, the physical network state is always the endpoint of the
executed prefix of the virtual trace, so the next concatenation starts
in its required state.  In the stateless case use the supplied union.
Thus physical network cost is at most virtual network cost $N_V$.
All admitted jobs have already arrived, and merging occurs only at a new
launch; no job joins a running batch.

Let $a_j$ be job $j$'s virtual service time and $s_j$ its physical start.
The next root completion clears the entire buffer, so
\[
 0\le s_j-a_j\le p_{\max}.
\]
This bound also holds when the virtual action occurs at a root completion,
in which case its delay is zero.  With $P_\Sigma=\sum_jp_j$ and
$G_V=\sum_j(a_j-r_j)$, physical cost is consequently at most
\begin{equation}
\label{eq:merge-add-upper}
 N_V+G_V+P_\Sigma+n p_{\max}
 \le N_V+G_V+(1+\rho)P_\Sigma.
\end{equation}
Every finite virtual trace and every finite processing length give a
finite physical completion.  There is no need to detect the final arrival.

Let $Q$ be the instantaneous optimum on the original requests.  Erasing
root occupation and processing from any feasible physical schedule leaves
its network actions, their state order, and its launch times as a feasible
instantaneous schedule.  Physical total flow is exactly $P_\Sigma$ plus
the total waiting to those launches.  Taking the infimum therefore gives
the objective-additive lower bound
\begin{equation}
\label{eq:merge-add-lower}
 P_\Sigma+Q\le\OPT_{\rm physical}.
\end{equation}
In particular, this uses one joint relaxation, rather than adding two
unrelated lower bounds on the physical optimum.  Virtual competitiveness
and~\eqref{eq:merge-add-upper} yield
\begin{align*}
 \ALG
 &\le \alpha Q+(1+\rho)P_\Sigma\\
 &\le \max\{\alpha,1+\rho\}(Q+P_\Sigma)
 \le \max\{\alpha,1+\rho\}\OPT_{\rm physical}.
\end{align*}
The empty input has cost zero.  The simulation only uses revealed jobs,
the virtual algorithm's current actions, and the next physical completion.
\end{proof}

\paragraph{Examples and scope.}
Line aggregation with delays (LAD) purchases a rooted prefix, joint
replenishment (JRP) pays a common order charge and the charges of the
included item types, and multi-level aggregation (MLA) purchases the union
of root-to-request paths.  Their supplied objects can be united without
increasing the sum of their costs.  Thus any verified deterministic
instantaneous total-wait guarantee for these models gives
Theorem~\ref{thm:merge-add} for their unbounded drop-line extensions.
For mobile service, the second realization property permits consecutive
zero-time walks to be replayed in order, retaining their endpoints and
movement costs.  This is a central-root launch model; it does not claim
the same bound for local production, positive travel time, or a requirement
to revisit each location after processing.  Additional per-job network
charges are permitted if the supplied unions or concatenations retain
their assigned jobs without increasing cost.  Capacity constraints that
prevent union are not covered by this theorem.

\subsection{A static relaxation for batch maxima}

For the sum of per-batch maxima, the same realization is useful after
processing is charged as part of a static service cost.  Unlike
Theorem~\ref{thm:merge-add}, the following result is stateless and uses
exact batch-cost optimization.

\begin{corollary}
\label{cor:merge-max-static}
Let $c$ be a finite, nonnegative, normalized, monotone subadditive cost
for stateless network service, with exact cost evaluation and realizable
minimizing actions.  Suppose the instantaneous maximum-waiting theorem
of~\cite{subadditive} is available for this cost-oracle model.  There is
a strict deterministic online algorithm with ratio $6-2/\rho$ for
\[
 \sum_{B}\left(c(B)+\max_{j\in B}(s_B-r_j+p_j)\right)
\]
on the unbounded drop-line processor.  It uses no advance processing-time
range.  A polynomial-time conclusion requires polynomial-time access to
the exact oracles required by that instantaneous theorem.
\end{corollary}

Here $c(B)$ denotes the minimum cost of an action for $B$.  If the
algorithm uses a supplied union instead, it pays that union's actual
cost; the proof bounds this actual expenditure as well.  It may also
invoke the assumed exact oracle to replace the union by a minimum-cost
action.

\begin{proof}
Set $P(\varnothing)=0$ and define
\[
 h(B)=c(B)+2P(B).
\]
The function $P$ is normalized, monotone, and subadditive, so $h$ has the
same properties.  Run the strict deterministic factor-two instantaneous
algorithm of~\cite{subadditive} with service cost $h$.  Its virtual value
is
\[
 V=\sum_b\bigl(c(B_b)+2P(B_b)+W_b\bigr),
 \qquad W_b=t_b-\min_{j\in B_b}r_j.
\]
Here processing lengths are revealed attributes used in the static
oracle; the virtual algorithm has no physical processor.

Merge virtual actions at root completions as in
Theorem~\ref{thm:merge-add}.  Let $\mathcal V_i$ be the virtual actions
merged into physical batch $i$, and let $P_i$ be that batch's maximum
processing time, with $P_0=0$.  Each buffered action is delayed by at
most $P_{i-1}$.  Consequently the physical maximum flow $M_i$ satisfies
\[
 M_i\le\max_{b\in\mathcal V_i}W_b+P_{i-1}+P_i.
\]
The supplied unions cost at most
$\sum_{b\in\mathcal V_i}c(B_b)$.  Since the nonempty families
$\mathcal V_i$ partition the virtual actions and
$P_i\le\sum_{b\in\mathcal V_i}P(B_b)$, summation gives
\[
 \ALG\le\sum_b\bigl(c(B_b)+W_b+2P(B_b)\bigr)=V.
\]

To compare its optimum with a physical optimum, fix a physical batch
launched at $s$, and write $P=P(B)$,
$W=s-\min_{j\in B}r_j$ and
$M=\max_{j\in B}(s-r_j+p_j)$.  The oldest job has processing time at
least $P/\rho$, while the longest job has flow at least $P$.  Hence
$M\ge W+P/\rho$ and $M\ge P$.  Serving the same batch instantaneously
at $s$ gives
\begin{align*}
 h(B)+W
 &\le c(B)+M+(2-1/\rho)P\\
 &\le (3-1/\rho)\bigl(c(B)+M\bigr).
\end{align*}
Thus $\OPT_h\le(3-1/\rho)\OPT_{\rm physical}$.  The virtual factor-two
bound now proves $\ALG\le V\le(6-2/\rho)\OPT_{\rm physical}$.
\end{proof}

For LAD on a rooted half-line, $c(B)$ is the farthest required distance
(possibly with a fixed nonempty-service charge).  For JRP it is the
common charge plus the charges of the represented item types, and for
MLA it is the weight of the unique rooted path union.  These costs satisfy
the hypotheses and have directly computable exact minimizing objects.

\subsection{Threshold prices and efficient Steiner purchases}\label{app:price}
\paragraph{Static purchases with small spread.}
Assume the static cost \(c\) satisfies the hypotheses of the paper's
instantaneous factor-two threshold theorem. For common processing time
\(p\), charge \(c(B)+p\) for every nonempty virtual purchase. Launch
all pending jobs when their oldest wait reaches this price. Every job
of the next batch arrived strictly after the preceding launch, since
arrivals at a timestamp are processed before a launch. Its timer allows
at least \(p\) units. Hence successive virtual launches are physically
feasible. The physical batch cost equals the virtual price plus oldest
wait, and deleting occupation gives a virtual optimum at most the
physical optimum. This proves factor two, tight over LAD, JRP and MLA
by their fixed-fee subinstances.

For arbitrary lengths assign each job the permanent mark
\(u_j=\max_{r_\ell\le r_j}p_\ell\), including the entire simultaneous
arrival group. Set \(h(B)=c(B)+\max_{j\in B}u_j\), with value zero
on the empty set. This normalized monotone subadditive function is
known on all currently available labels. The next batch's timer allows
at least the preceding duration, by the same release-order argument.
Physical cost is at most virtual cost. On any physical comparator
batch of cost \(c+M\), the virtual cost is at most
\(c+W+p_{\max}\le\rho(c+M)\), since \(M\ge W+p_{\min}\).
This gives \(2\rho\) without an announced range.

\paragraph{Tuning the static threshold.}
Let \(D(B)\) be a computable nonnegative normalized monotone subadditive
price, with a service object of cost at most \(D(B)\) for every batch.
Suppose supplied objects can be combined at no extra cost and
\(D(B)\le\gamma c(B)\), where \(c\) is the exact physical minimum.
For \(1\le\lambda\le2\), use the virtual all-pending threshold with
\(h=D+\lambda P\). This virtual problem permits buying any set at
its declared price. Its objective \(V\) satisfies
\[
 V=\sum_b (h_b+W_b)=2\sum_b(D_b+\lambda P_b)
       \le2\operatorname{OPT}_h.
\]
Equality at launch holds since the oldest wait grows continuously
between arrival events and arrivals can only increase the threshold.
Pool nonempty virtual events during each physical occupation. The
maximum flow in a physical group is at most its largest virtual wait,
plus its own and its predecessor's durations. The proof in the
aggregation appendix then gives
\[
 \mathrm{ALG}\le\sum_b(D_b+W_b+2P_b)
 =\sum_b(2D_b+(\lambda+2)P_b)
 \le (1+2/\lambda)\operatorname{OPT}_h .
\]
An optimum physical batch of cost \(c+M\) is available virtually at
the same instant. Its virtual cost satisfies
\[
 D+\lambda P+W\le\gamma c+(1+\lambda-1/\rho)M.
\]
Together these inequalities give the ratio
\((1+2/\lambda)\max\{\gamma,1+\lambda-1/\rho\}\).
With \(D=c\) and \(\lambda=\sqrt2\), this is
\((1+\sqrt2)(1+\sqrt2-1/\rho)\). This replay policy and the
historical-mark policy above are distinct algorithms.

\paragraph{Rooted Steiner purchases in polynomial time.}
In a known finite undirected graph with nonnegative rational edge
costs and reachable terminals, define \(L(B)\) by the undirected
Steiner cut relaxation. Minimize \(\sum_e w_e x_e\) over \(x\ge0\)
subject to \(x(\delta(S))\ge1\) whenever \(S\) excludes the root
and contains a terminal of \(B\). Set \(D(B)=2L(B)\).
Adding terminals adds constraints. Adding two feasible vectors yields
a feasible vector for the union. Thus \(D\) is monotone and
subadditive. A physical tree gives a feasible indicator vector, so
\(D(B)\le2c(B)\). The primal-dual constrained-forest algorithm of
Goemans and Williamson~\cite[Theorem~2.4]{gw} with
Steiner cuts returns an actual tree of cost at most \(2L(B)\);
the approximation is certified against a feasible LP dual. The union
of the supplied rooted trees can be pruned without increasing its cost.
The value \(L\) can be computed exactly using root-to-terminal
minimum-cut separation, or a polynomial-size multicommodity flow
formulation with a separate unit flow from the root to each terminal
subject individually to the common edge capacities. A polynomial number
of threshold updates and tree constructions suffices. With rational
release and processing data, \(\lambda=2\) uses rational arithmetic;
the fixed constant \(\lambda=\sqrt2\) only requires arithmetic and
comparisons in a fixed quadratic extension.

The preceding bound with \(\gamma=2,\lambda=2\) gives precisely
\(6-2/\rho\), now in polynomial time. With \(\lambda=\sqrt2\) it
gives a uniform \(3+2\sqrt2\) bound. The spread-sensitive expression
uses the maximum with \(2\) above; it need not coincide with the
exact-cost tree expression at small spreads.

\subsection{A direct timer for line aggregation}\label{app:lad-paired}
Consider a rooted half-line with job distances \(x_j\ge0\) and
processing lengths \(p_j>0\). For every nonempty subset \(B\), set
\[
 X(B)=\max_{j\in B}x_j,\qquad d(B)=\max_{j\in B}(x_j+p_j).
\]
A physical batch pays \(X(B)+M(B,s)\). For a nonempty instance define
\[
 \eta=\max_{\varnothing\ne B\subseteq J}\frac{X(B)+P(B)}{d(B)}.
\]
We have \(1\le\eta\le2\). The algorithm uses the paired values
\(x_j+p_j\) as jobs arrive and needs no bound on \(\eta\).

\begin{theorem}\label{thm:lad-paired-timer}
Launch all pending jobs \(B\) as soon as the processor is free and
their oldest waiting time is at least \(d(B)\). This deterministic
algorithm has cost at most \((3+2\eta)\OPT\) for network cost plus
the sum of batch maximum flow times. In particular its competitive
ratio is exactly five on the class of instances admitting a single
job order nondecreasing in both \(x_j\) and \(p_j\).
\end{theorem}
\begin{proof}
For each job draw \(I_j=[r_j-x_j-p_j,r_j]\) and write \(U=\bigcup_j I_j\).
Any comparator batch \(B\) launched at \(s\) covers all its job intervals
by the interval from their leftmost endpoint to \(s\), of length
\(\max_{j\in B}(s-r_j+p_j+x_j)\le M(B,s)+X(B)\).
Consequently \(|U|\le\OPT\).

For the \(i\)th algorithmic batch write \(b_i,s_i,P_i,d_i,w_i\) for
its oldest release, start, duration, threshold, and oldest wait.
Put \(E_i=s_i+P_i\), with \(E_0=0\). The threshold can only increase
between the first arrival and launch. The timer thus gives
\(s_i=\max\{b_i+d_i,E_{i-1}\}\). Write
\(\delta_i=w_i-d_i\ge0\). We have \(\delta_1=0\), and
\(\delta_i\le(d_{i-1}-d_i)_+\) for \(i\ge2\), since
\(b_i>s_{i-1}\) and \(P_{i-1}\le d_{i-1}\). Its physical cost
is at most \(w_i+X(B_i)+P_i\le (w_i+d_i)+(\eta-1)d_i\).

Assign batch \(i\) to the component of \(U\) containing the interval
of a job attaining \(d_i\), and let \(v_i\) be that job's release.
Since \(v_i\le s_i<b_{i+1}\le v_{i+1}\), each used component receives
a consecutive block. Within one such component of length \(H\), number
its batches locally from 1 to \(k\). Every \(d_i\le H\).
For \(k\ge2\), the first representative interval ends at \(v_1\);
the waiting intervals \([b_i,s_i]\) for \(2\le i<k\) follow it,
are disjoint, and end by \(v_k\). All lie in the component, so
\(d_1+\sum_{i=2}^{k-1}w_i\le H\).
For the first and interior batches, \(w_1+d_1=2d_1+\delta_1\)
and \(w_i+d_i\le2w_i\). Their sum is at most \(2H+\delta_1\).
For the last batch \(w_k+d_k\le
d_k+\max\{d_k,d_{k-1}\}\le2H\). With a single batch the bound is
just \(2H+\delta_1\). Here the local \(\delta_1\) may come from the
last batch of the preceding used component, whose length bounds it.
Each used component has at most one successor, and the first has zero
initial excess. Summing gives \(\sum_i(w_i+d_i)\le5|U|\).
The same packing and \(d_i\le w_i\) give \(\sum_i d_i\le2H\)
within each component: the first and interior thresholds fit in \(H\)
and the last threshold is at most \(H\). Thus the additional cost
is at most \(2(\eta-1)|U|\), proving the claimed ratio.

In a common nondecreasing order the last job of every batch attains
both maxima, so \(\eta=1\). On inputs with constant positive distance
\(x_j=K\) the algorithm and its comparator specialize exactly to TCP.
Theorem~\ref{thm:max-timer-tight} gives the matching lower bound for
this timer. The empty input incurs zero cost.
\end{proof}

\subsection{The fixed-fee specialization}\label{app:static}
\begin{corollary}\label{prop:fixed-spread-virtual}
For TCP, the static-cost transfer gives a deterministic ratio of
\(6-2/\rho\) for the batch-maximum objective.
\end{corollary}
\begin{proof}
Apply Corollary~\ref{cor:merge-max-static} with \(c(B)=K\) for nonempty
\(B\). The virtual cost is \(K+2P(B)\) per service plus maximum waiting.
\end{proof}
The direct pending timer in Section~\ref{sec:tcp-main} is stronger for
TCP. The static-cost proof is useful on a tree because virtual services
can charge distinct sets of purchased edges.

\section{Ordered walk pooling for mobile service}
\label{app:osd}

We consider the drop-line version of online service with delay in a
metric space.  A courier starts at a prescribed point.  A service batch
selects pending jobs and a walk visiting their locations, starting at
the courier's current position.  At launch time $s$ this walk is
traversed instantaneously; selected job $j$ completes at $s+p_j$, and
the root remains occupied until $s+\max_j p_j$.  The next service walk
cannot start before that time.  The courier remains at the walk's final
point.  Processing does not require the courier to remain at a job's
location.  Service is elective: traversing a location need not assign
other jobs there to the current batch.  There is no prescribed final
courier position.  Movement costs the length of the walk.

The proof retains this moving state explicitly.  It does not replace a
walk by a static set cost.  Both the instantaneous request problem and
the physical problem use the same elective service convention.
Assume that an instantaneous request algorithm produces feasible walks,
serves every request on finite
inputs, and provides a finite description of the walk prefix between
successive nonempty service events.  Processing times are known upon
arrival, but the virtual request algorithm runs only on the original
request arrivals and locations.

\subsection{The pooling construction}

Assign every job to the first virtual event that serves its request.
Enumerate the resulting nonempty events chronologically as
$h=1,\ldots,m$, breaking ties in the virtual itinerary's order.  Event
$h$ has time $a_h$ and assigned identity set $B_h$.  The sets $B_h$
partition the jobs.  A zero-length walk can be a nonempty event.
Bundle all intervening movements without assigned jobs into the walk
preceding the next nonempty event.  Denote the resulting walk by $Q_h$.
Its start is the endpoint of $Q_{h-1}$, with the initial position used
for $Q_1$.  Optional trailing movements are omitted.  Consequently
\[
  \sum_h\operatorname{len}(Q_h)\leq N_V,
\]
where $N_V$ is the virtual algorithm's total movement.

The physical algorithm continues the virtual simulation while the root
is occupied.  Whenever the root is free and at least one nonempty
virtual event has not yet been launched physically, concatenate all
currently formed, unlaunched walks $Q_h$ in their original order, and
start one physical batch containing the union of their assigned jobs.
If no such event is available, wait for the next one.  All virtual
events at a physical timestamp, including their assigned arrivals, are
processed before a new physical launch decision.

Thus each physical batch $i$ corresponds to a consecutive block
$\mathcal H_i$ of virtual event indices.  Let $s_i$ be its launch,
$P_i$ its longest processing time, and $E_i=s_i+P_i$ its occupied
interval's end.  Set $P_0=0$.  Concatenation is state-consistent:
after a physical launch the courier is at the virtual endpoint of the
last consumed walk.  The next buffered walk starts at exactly that
point.  The physical courier may stay there while its root is occupied,
even though the independent virtual courier keeps moving.  Every
consumed leg is paid only once, so physical movement is at most $N_V$.

For $i\geq2$ and $h\in\mathcal H_i$, the event was formed after the
previous physical launch; otherwise it would have been included then.
In particular,
\begin{equation}\label{eq:osd-pooling-time}
  a_h\geq s_{i-1},\qquad
  0\leq s_i-a_h\leq P_{i-1}.
\end{equation}
For the first batch the latter delay is zero.  Indeed, if an event is
buffered during the preceding occupied interval, all such events launch
at its end.  If none is buffered, the next event launches immediately
when it occurs.  These assertions remain valid at simultaneous
completion and arrival times under the stated event convention.
Every job is available when launched, and every finite virtual prefix
is eventually executed without requiring an end-of-input signal.

\subsection{Sum of batch maximum flow times}

Suppose the virtual algorithm has competitive ratio $\alpha$ against
the elective instantaneous optimum for movement plus the sum of
per-event maximum waiting times.  Write
\[
  W_h=\max_{j\in B_h}(a_h-r_j),\qquad
  W_V=\sum_h W_h,\qquad V=N_V+W_V.
\]
The physical objective is movement plus
$\sum_i M_i$, where
$M_i=\max_{j\in B_i^{\rm real}}(s_i-r_j+p_j)$.

\begin{lemma}\label{lem:osd-real-flow}
The pooling construction satisfies
\[
  \mathrm{ALG}\leq N_V+W_V+2\sum_i P_i.
\]
\end{lemma}

\begin{proof}
For $j\in B_h$ with $h\in\mathcal H_i$,
\[
  s_i-r_j+p_j
   =(a_h-r_j)+(s_i-a_h)+p_j
   \leq W_h+P_{i-1}+P_i.
\]
The same inequality holds for $i=1$ with $P_0=0$.
Nonnegativity of the virtual waiting charges gives
\[
  M_i\leq\sum_{h\in\mathcal H_i}W_h+P_{i-1}+P_i.
\]
Summing uses each virtual event once and charges a preceding processing
length at most once.  Together with the movement bound this proves
the assertion.
\end{proof}

\begin{lemma}\label{lem:osd-duration-packing}
For every feasible physical comparison schedule with total batch
maximum flow $F^*$,
\[
  \sum_i P_i\leq3F^*+W_V.
\]
\end{lemma}

\begin{proof}
For each comparison batch $H$ started at $t_H$, define
\[
  I_H=[\min_{j\in H}(r_j-p_j),t_H].
\]
Its length is exactly the comparison batch's maximum flow
$M_H^*=\max_{j\in H}(t_H-r_j+p_j)$, so
$\sum_H|I_H|=F^*$.  For each virtual event define
\[
  J_h=[\min_{j\in B_h}r_j,a_h],\qquad |J_h|=W_h.
\]
Let
\[
  U=\left(\bigcup_H I_H\right)\cup
       \left(\bigcup_h J_h\right).
\]
This is a finite union of closed intervals, possibly extending to
negative times, and
\begin{equation}\label{eq:osd-union-length}
  |U|\leq F^*+W_V.
\end{equation}

Choose a longest job $j_i$ in each real batch, so $p_{j_i}=P_i$.
Let $h_i$ be its virtual event and let $H_i^*$ be its comparison batch.
The interval
\[
  [r_{j_i}-p_{j_i},a_{h_i}]
\]
is contained in $I_{H_i^*}\cup J_{h_i}$: the first interval covers
its part up to $r_{j_i}$ and the second covers the remaining part.
Both intervals contain $r_{j_i}$, so their union is connected.
Assign real batch $i$ to this connected component of $U$.  In
particular the entire interval $I_{H_i^*}$ belongs to the same component.

Fix one used component $[L,R]$, and let $\mathcal A$ be the set of real
batch indices assigned to it. The argument uses only their order in
the full physical schedule.
Let $k=\max\mathcal A$.  If $k\geq2$,
\eqref{eq:osd-pooling-time} gives
\[
  s_{k-1}\leq a_{h_k}\leq R.
\]
For every assigned index $i\leq k-2$, its occupied interval starts
at $s_i\geq a_{h_i}\geq L$ and ends at
\[
  E_i\leq s_{i+1}\leq s_{k-1}\leq R.
\]
The occupied intervals are disjoint.  Their total length is therefore
at most $R-L$.  The only assigned indices not accounted for this way
are in $\{k-1,k\}$, at most two indices.  For $k=1$ there is simply
one such exceptional batch and no packed interval.

For each exceptional index $i$,
\[
  P_i=p_{j_i}\leq M_{H_i^*}^*,
\]
since the comparison batch cannot start before $r_{j_i}$.  Moreover
$I_{H_i^*}$ lies wholly in this same component.  At most two exceptions
belong to any component, and a comparison interval cannot belong to
two components.  Hence each comparison batch is charged at most twice
over the entire argument, even if it contains many selected jobs.
The exceptional lengths sum to at most $2F^*$.  Summing the packed
lengths over components and using~\eqref{eq:osd-union-length} yields
\[
  \sum_i P_i\leq |U|+2F^*\leq3F^*+W_V.
\]
\end{proof}

\begin{theorem}\label{thm:osd-max-transfer}
An $\alpha$-competitive instantaneous request algorithm for movement
plus the sum of per-event maximum waiting times yields a
$(3\alpha+6)$-competitive algorithm for elective drop-line OSD with
movement plus the sum of batch maximum individual flow times.
The construction requires neither a processing-time ratio nor a final
job count.
\end{theorem}

\begin{proof}
Let a physical optimum have movement $N^*$ and maximum-flow sum $F^*$.
The two preceding lemmas give
\[
  \mathrm{ALG}\leq N_V+3W_V+6F^*\leq3V+6F^*.
\]
Erase the occupation intervals of the comparison schedule, but keep
its service walks, their order, and their start times.  This is a
feasible instantaneous request schedule with the same courier states
and movement.  Its waiting charge for a batch is at most that batch's
maximum individual flow.  Thus the instantaneous optimum is at most
$N^*+F^*$ and $V\leq\alpha(N^*+F^*)$.  Consequently
\[
  \mathrm{ALG}
    \leq3\alpha N^*+(3\alpha+6)F^*
    \leq(3\alpha+6)\mathrm{OPT}.
\]
The empty input has zero cost.  Every nonempty real batch has positive
occupation length because the original processing times are positive;
zero-length virtual walks cause no exception.  Pure movements are
handled by the ordered bundling above, and the algorithm serves only
its explicitly selected identities.  In particular it does not rely
on automatic inclusion of a newly arrived job in a running batch.
\end{proof}

The bounds are pathwise in the virtual schedule.  Therefore the same
ratio holds in expectation when the request algorithm has the stated
expected guarantee against a fixed oblivious input.  A request bound
with additive constant $\beta$ instead propagates an additive
$3\beta$ in this maximum-flow theorem.  The polynomial implementation
of an underlying request algorithm is preserved: there are at most
one nonempty event and one real batch per input job, and walk prefixes
are concatenated without solving a new spatial optimization problem.

\subsection{The additive-flow walk application}

For completeness, ordered pooling also supplies the moving-state
analogue of Theorem~\ref{thm:merge-add}.  Let an instantaneous request
OSD algorithm be $\alpha$-competitive for movement plus the sum of
individual waiting times.  Put $P=\sum_jp_j$ and
$\rho=p_{\max}/p_{\min}$.  Use the same pooling construction, and let
$a_j$ be job $j$'s virtual service time.  Its real start $s_j$ satisfies
$0\leq s_j-a_j\leq p_{\max}$ by
\eqref{eq:osd-pooling-time}.  Therefore, if $G_V=\sum_j(a_j-r_j)$,
\[
  \sum_j(s_j-r_j+p_j)
    \leq G_V+P+n p_{\max}
    \leq G_V+(1+\rho)P.
\]
Movement again costs at most $N_V$, by state-consistent ordered
concatenation.  These are exactly the two estimates needed in the
proof of Theorem~\ref{thm:merge-add}; no static set-cost assumption for
mobile walks is required here.

Indeed, let $Q^*$ be the instantaneous additive request optimum.
Erasing occupation from any physical comparison schedule and serving
its batches at their original starts subtracts exactly $P$ from its
objective, without changing movement.  Hence
$Q^*+P\leq\mathrm{OPT}$.  The resulting guarantee is
\[
  \mathrm{ALG}_{\rm add}
    \leq\alpha Q^*+(1+\rho)P
    \leq\max\{\alpha,1+\rho\}\,\mathrm{OPT}.
\]
The algorithm need not know $\rho$.  This application uses linear
individual flow, elective service, unbounded batches, and instantaneous
walks; the moving endpoints are preserved throughout the construction.

\section{Ephemeral facilities with a shared batch processor}
\label{app:facility}

Let \(\mathcal F\) be a known set of \(k\ge1\) candidate sites.
Opening site \(f\) for one batch costs \(a_f\ge0\), and assigning job
\(j\) to it costs \(\ell_{fj}\ge0\).  These costs are static, and the
connection costs are known when the job arrives.  Write
\[
 g_f(B)=a_f+\sum_{j\in B}\ell_{fj}\qquad(B\ne\varnothing).
\]
A physical batch chooses exactly one site and starts an elective subset
of the released, unstarted jobs.  It pays \(g_f(B)\), completes job
\(j\) at \(s+p_j\), and occupies the single shared processor until
\(s+P(B)\).  Openings are ephemeral: a later batch at the same site
pays again.  Processing times are positive and known on arrival.
There is no capacity bound, joining, or preemption.  The two objectives
are the sum of service costs plus, respectively, total job flow or the
sum of the maximum job flow within each physical batch.  All statements
below concern nonempty finite inputs; the empty input costs zero.

\subsection{A common replay construction}

Consider a virtual stream of instantaneous star events.  Each event
assigns its jobs to one site, and every job appears in exactly one
event.  When the physical processor is idle, start a round using the
next virtual event.  Freeze the round's membership at that instant,
pool its jobs by assigned site, and process its nonempty site batches
in a fixed order.  Events occurring during the round are buffered for
the next round.  When a round ends, freeze all buffered events and
start their pooled round immediately, if the buffer is nonempty;
otherwise wait for the next virtual event.  Simultaneous virtual
events are collected before freezing a new round.  Virtual events
arriving during a round never join any of its physical batches.

This is an online feasible construction: every pooled job has already
arrived, each round contains at most \(k\) sequential physical batches,
and those batches use the common processor without overlap.  Pooling
events at a fixed site pays its opening only once and preserves every
connection charge.  Consequently physical service cost is at most
virtual service cost.  The virtual simulation is independent of the
physical state; its events need not wait for physical completions.

\begin{theorem}
\label{thm:facility-total-transfer}
Suppose the instantaneous ephemeral-star problem with total waiting
has a strict deterministic \(\alpha\)-competitive online algorithm,
where \(\alpha\ge1\).  Then the physical total-flow problem has a
strict deterministic online algorithm with ratio
\[
 \max\{\alpha,\,1+(2k-1)\rho\}.
\]
The resulting algorithm needs no processing-time spread bound.
\end{theorem}

\begin{proof}
Run that virtual algorithm on the original releases and replay its
events in rounds.  Let \(t_j\) be job \(j\)'s virtual service time.
A round lasts at most \(kp_{\max}\), so the residual preceding-round
delay of this job is at most \(kp_{\max}\).  Its site batch is preceded
by at most \(k-1\) other batches in the new round.  Thus its physical
flow satisfies
\[
 C_j-r_j\le (t_j-r_j)+(2k-1)p_{\max}+p_j.
\]
For \(Q=\sum_jp_j\), virtual service cost \(N_V\), and virtual total
waiting \(G_V\), the replay therefore obeys
\begin{equation}
 \label{eq:facility-total-replay}
 \ALG\le N_V+G_V+Q+(2k-1)np_{\max}.
\end{equation}

Fix a physical optimum with service cost \(N^*\) and total flow
\(F^*\).  Erase occupancy and place each of its virtual star events
at the corresponding physical \emph{start} time.  This is a feasible
instantaneous schedule: all jobs of a physical batch were released
before its start.  Its total waiting is exactly \(F^*-Q\).
The virtual competitive guarantee hence gives
\[
 N_V+G_V\le\alpha(N^*+F^*-Q).
\]
This subtraction of mandatory processing is essential to the stated
maximum of two factors.  Since \(np_{\max}\le\rho Q\), putting
\(\beta=1+(2k-1)\rho\) in
\eqref{eq:facility-total-replay} yields
\[
 \ALG\le\alpha(N^*+F^*)+(\beta-\alpha)Q
       \le\max\{\alpha,\beta\}(N^*+F^*),
\]
using \(0\le Q\le F^*\le N^*+F^*\).  The replay itself uses only
revealed jobs and virtual assignments; \(p_{\max}\) and \(\rho\)
occur solely in its analysis.
\end{proof}

\subsection{Maximum flow charged separately to each star}

It is important to retain the atomic-star convention.  One virtual
event below may use several sites, but each of its realized physical
site batches pays its own maximum flow.  Define the normalized static
set function
\[
 h(B)=\min_{\substack{B=\dot\bigcup_{f\in\mathcal F}B_f}}
 \sum_{f:B_f\ne\varnothing}
   \left(a_f+\sum_{j\in B_f}\ell_{fj}+2kP(B_f)\right),
 \qquad h(\varnothing)=0.
\]
It is monotone: restrict an assignment after deleting jobs and discard
any unused site.  It is subadditive: combine two assignments site by
site, paying an opening at most once and using
\(P(A\cup B)\le P(A)+P(B)\).  These arguments also apply to disjoint
labelled batches with identical spatial attributes.

\begin{lemma}
\label{lem:facility-static-oracle}
For fixed \(k\), an optimal assignment attaining \(h(B)\) can be
computed with \((|B|+1)^k\) cutoff vectors and polynomial work per
vector.  The running time is independent of the numerical spread.
\end{lemma}

\begin{proof}
For each site enumerate a cutoff from \(\{0\}\cup\{p_j:j\in B\}\),
where zero means closed.  A positive cutoff \(u_f\) incurs
\(a_f+2ku_f\).  Assign each job to its cheapest-connection eligible
site among those with \(u_f\ge p_j\); reject a vector if some job has
no eligible site.  Evaluate the resulting assignment with its actual
nonempty sites and maximum processing times, discarding unused sites.
Every evaluated assignment is feasible.  For an optimal assignment,
its site maxima form an enumerated vector.  At that vector the
cheapest eligible connections and subsequent removal of unused sites
cost no more than the optimum assignment.  The minimum evaluated
assignment is therefore optimal.  This enumeration takes
\(O(k|B|(|B|+1)^k)\) arithmetic operations; rational input costs have
polynomial encoding length.  It is polynomial for fixed \(k\), not
a polynomial-time assertion for unrestricted \(k\).
\end{proof}

\begin{theorem}
\label{thm:facility-max-transfer}
For fixed \(k\), the physical ephemeral-facility problem with a sum of
per-batch maximum flows admits a deterministic online algorithm with
strict competitive ratio
\[
 k(6-2/\rho).
\]
It uses exact static assignments and needs no advance knowledge of
\(n\), \(p_{\min}\), \(p_{\max}\), or \(\rho\).
\end{theorem}

\begin{proof}
Apply the deterministic \(2\)-competitive instantaneous batching
algorithm of Lu, Ren, and Liu~\cite{subadditive} to service cost
\(h/k\) and per-event maximum waiting.  Multiplying its objective by
\(k\), its virtual cost is
\[
 V=\sum_b\bigl(h(B_b)+kW_b\bigr),\qquad
 W_b=t_b-\min_{j\in B_b}r_j,
\]
and it is at most twice the optimum for this virtual objective.
Retain an exact minimizing site assignment for each virtual event,
and replay its assigned stars using the preceding rounds.

Let \(\mathcal V_i\) be the virtual events frozen into round \(i\),
let \(P_{if}\) be its maximum processing time at used site \(f\), and
write \(D_i=\sum_fP_{if}\) for the round's duration, with \(D_0=0\).
For every physical star in that round, its maximum job flow is at most
\[
 \max_{b\in\mathcal V_i}W_b+D_{i-1}+D_i.
\]
Indeed, a buffered event waits at most \(D_{i-1}\) until this round
starts.  Within this round, the preceding site batches together with
the current job's processing consume at most \(D_i\).  There are at
most \(k\) nonempty stars in a round.  If \(B_{bf}\) is the assigned
part of virtual event \(b\), pooling gives
\(D_i\le\sum_{b\in\mathcal V_i,f}P(B_{bf})\).  Thus
\[
 \sum_{\text{physical stars}}\max_j F_j
 \le k\sum_bW_b+2k\sum_{b,f}P(B_{bf}).
\]
Together with nonincreasing service cost, this proves \(\ALG\le V\).
The two duration charges explicitly pay both preceding-round blocking
and all processing within the new round.

Now consider an atomic optimum star \(B\) launched at \(s\).  Write
\(P=P(B)\), \(W=s-\min_{j\in B}r_j\),
\(M=\max_{j\in B}(s+p_j-r_j)\), and \(g\) for its service cost.
Using its one site is an admissible assignment for \(h\).  Moreover,
\(W+P/\rho\le M\) and \(P\le M\).  Hence its instantaneous
virtual event at time \(s\) has cost at most
\[
 \begin{aligned}
 h(B)+kW
 &\le g+2kP+kW\\
 &\le g+kM+(2k-k/\rho)P\\
 &\le g+k(3-1/\rho)M\\
 &\le k(3-1/\rho)(g+M).
 \end{aligned}
\]
The resulting virtual schedule is feasible after erasing occupation.
Therefore \(\OPT_V\le k(3-1/\rho)\OPT\), and
\(\ALG\le V\le2\OPT_V\) proves the result.  The cited instantaneous
algorithm is used with its value-and-realization oracle contract;
Lemma~\ref{lem:facility-static-oracle} supplies that oracle for fixed
\(k\).  All physical actions involve nonempty sets of already
released jobs, so a finite input is completed without an end signal.
\end{proof}

\paragraph{A uniform constant for unrestricted lengths.}
More generally use \(\lambda kP(B_f)\) in each site's virtual price
in place of \(2kP(B_f)\), for \(1\le\lambda\le2\). Denote the new
minimum assignment price by \(h_\lambda\). At each virtual launch
\(kW=h_\lambda\). In the minimizing assignment write \(g\) for the
sum of spatial charges and \(S\) for the sum of site durations. The
same round analysis charges at most \(g+kW+2kS\) per event after
summation. Since \(g+kW+2kS=2g+(\lambda+2)kS\le
(1+2/\lambda)h_\lambda\), and \(\sum h_\lambda\le\OPT_V\),
it suffices to bound the virtual optimum. An offline star induces
virtual price at most \(g+k(1+\lambda-1/\rho)M\), by the preceding
comparison. Choosing \(\lambda=\sqrt2\) therefore yields the uniform
bound \(k(3+2\sqrt2)\) without advance knowledge of the spread.

\subsection{Why a single atomic star cannot be merged freely}

Even with metric connection costs, the cheapest \emph{one-star} cost
need not be subadditive.  Take two candidate sites at distance \(L\),
each with opening cost one, and one job at each site.  Either singleton
has one-star cost one, whereas their union has one-star cost \(1+L\).
The ratio to the sum of singleton costs is unbounded as \(L\) grows.
The virtual function \(h\) therefore permits several assigned stars,
and the replay pays a separate maximum flow for each physical star;
it does not identify their union with a single cheap atomic service.

Finally, one site with opening cost \(K\) and zero connections is
exactly TCP-BP.  The total-flow lower bound of
Theorem~\ref{thm:total}, the equal-length maximum-flow lower bound of
Theorem~\ref{thm:equal-two}, and the unrestricted deterministic lower
bound of Theorem~\ref{thm:max-unrestricted-lower} consequently apply
already to this one-site facility subcase. Several distinct short-job
sites can each incur the delay from the same occupation.

\subsection{A lower bound from one long job and several sites}

\begin{theorem}\label{thm:facility-count-lower}
For the sum of batch maximum flows plus service cost, every deterministic
online algorithm has competitive ratio at least \((k+1)/2\) over inputs
with \(k\) facilities and unrestricted spread. This holds on uniform
metrics with unit opening fees and at most two processing lengths.
At a prescribed spread \(\rho\ge1\), the same construction gives a lower
bound of \((1+\min\{k,\lfloor\sqrt{\rho}\rfloor\})/4\).
\end{theorem}

\begin{proof}
Fix short length \(p>0\), long length \(P\ge p\), and an integer
\(1\le m\le k\). The clients and facilities are colocated on the
same \(k\) points; distinct points have distance \(D=(k+2)P\).
Release one long job at time zero. The algorithm must eventually start
it at some finite \(T\) if no further jobs arrive. At \(R=T+p/2\),
release one job of length \(p\) at each of \(m\) distinct sites,
including the long job's site. The launch at \(T\) has already fixed
its membership. For a deterministic algorithm the construction
specifies a finite input.

A nonlocal assignment costs at least \(D\ge(m+1)P\). If all assignments
are local, the short jobs require \(m\) distinct new batches. Each
finishes no earlier than \(T+P+p\), so each contributes flow of at
least \(P+p/2\). The long job's batch contributes at least \(P\).
In either case \(\ALG\ge(m+1)P\).

Offline uses only local assignments. If \(R\ge P\), process the long
job at zero and the short jobs consecutively starting at \(R\).
The short flows sum to \(p\,m(m+1)/2\). If \(R<P\), process the
short jobs in the same way and then launch the long job at \(R+mp\).
Its flow is \(R+mp+P<2P+mp\). Both schedules pay \(m+1\) opening
fees, giving in all cases
\[
    \OPT\le m+1+2P+mp+p\,m(m+1)/2 .
\]
For unrestricted spread set \(m=k\), fix \(p=1\), and send
\(P\) to infinity to obtain \((k+1)/2\).
For prescribed \(\rho\) set \(P=\rho p\) and
\(m=\min\{k,\lfloor\sqrt{\rho}\rfloor\}\). Both \(mp\) and
\(p\,m(m+1)/2\) are at most \(P\). Let \(p\) tend to infinity
to absorb the unit fees and obtain \((m+1)/4\).
\end{proof}

Together with Theorem~\ref{thm:facility-max-transfer}, this gives
order \(k\) for the optimal deterministic ratio uniform over the
processing lengths. The optimal dependence on \(k\) at small
prescribed spread remains undetermined by these bounds.

\section{Further related work}\label{app:related}

\paragraph{Acknowledgment and the cost of a batch.}
Dooly, Goldman, and Scott~\cite{tcp} studied both the sum of packet
waiting times and the sum of the largest waiting times within individual
acknowledgments. They gave offline dynamic programs and deterministic
factor-two guarantees without lookahead. We use their accumulated-wait
threshold as a virtual policy for total flow. Each virtual decision is
instantaneous even when the physical root is busy. The resulting
buffering delay is precisely where the ratio of processing lengths
enters the analysis. Albers and Bals~\cite{albers} instead charge one
largest waiting time over the entire input. Bhore, Paw{\l}owski, and
Umboh~\cite{batchaware} distinguish aggregating delays within a batch
from aggregating costs across batches. Our second objective retains
one charge for each batch an algorithm actually launches.

Frederiksen, Larsen, Noga, and Uthaisombut~\cite{logp} already model
a transmission that blocks later transmissions. Their LogP model
prescribes a fixed minimum gap. Its principal objective measures the
length of the union of the intervals during which at least one message
remains undelivered. They also show that charging each message only
until its send time, together with a packet fee, admits no bounded
competitive ratio under a fixed gap. Their deterministic adversary
releases a burst just after a transmission. We use this adversarial
idea in Appendix~\ref{sec:total}. Here the gap depends on the chosen
batch, and each new job also incurs its own processing time before
completion. That unavoidable cost gives a finite ratio at each fixed
spread, and the burst determines its exact dependence on the spread.

\paragraph{Aggregation and mobile service.}
MLA buys a rooted subtree containing the requests served~\cite{mla}.
For per-batch maximum waiting we use the static subadditive
service-cost result~\cite{subadditive}. In OSD the starting location for a service
depends on the preceding walk~\cite{osd,osdmax}. TCP-BP carries a
different piece of information between decisions: the earliest time at
which the root becomes free. It can matter even when all jobs require
the same network resource. Lu, Ren, and Liu~\cite{subadditive} extend
the instantaneous maximum-wait guarantee to static subadditive service
costs with exact batch optimization. Appendix~\ref{app:aggregation} explains
what can be retained by charging processing as a static service cost;
the direct timer of Appendix~\ref{sec:max-upper} yields the stronger
bound for the occupied root.

\paragraph{Online and offline drop-line scheduling.}
Tian, Wang, Fu, and Yuan~\cite{online-drop} allow online releases,
unbounded batch size, joint starts, individual completion at \(s+p_j\),
and machine occupation until the longest member finishes. Restricted
to one machine, this is the same service mechanism as TCP-BP. Their
objective minimizes \(\max_j(C_j+q_j)\), where \(q_j\) is delivery
time after processing. Even with zero delivery times this measures
the last completion; moving one early batch or opening an extra batch
can change either of our fee-plus-flow objectives without changing
that completion. Thus we use the existing scheduling model and analyze
a different cost of its launch decisions.

Gao, Yuan, and Wei~\cite{dropline} study offline drop-line jobs with
release dates and regular objectives of the form \(\sum_j f_j(C_j)\)
or \(\max_j f_j(C_j)\). For the general classes they give
pseudo-polynomial algorithms and polynomial algorithms for a constant
number of processing times or release dates. With \(K=0\), minimizing
total flow and minimizing total completion time have the same
minimizers, since \(\sum_j r_j\) is fixed. Their numerical values differ,
so a multiplicative approximation for completion time need not preserve
its ratio after that subtraction. With \(K>0\) the schedule also chooses
how many fees to pay, and the grouping itself determines each batch
maximum. Class-level hardness for a general regular objective does not
classify either of these fixed-fee specializations. Our offline results
in Appendix~\ref{sec:offline} are direct guarantees for these objectives.

Ji, He, and Cheng~\cite{batchdelivery} charge total weighted flow time
and a cost depending on the number of deliveries. Their jobs are
available together, undergo single-machine processing, and receive a
common delivery date in each shipment. A shipment's delivery date is
the completion time of its last processed job. In TCP-BP each paid
launch starts its members simultaneously, and individual completion
times can differ within that batch. The known batch-delivery hardness
result therefore concerns a different feasible schedule and objective.

\paragraph{Flow objectives in online batching.}
Gfeller et al.~\cite{gfeller2009} study online single-machine batching
with setup time, motivated by logging. A batch processes its jobs
sequentially and makes them available together. Its occupied time is
the setup time plus the sum of its processing lengths. In TCP-BP,
the occupied time is the maximum processing length, each member
finishes individually, and the opening fee enters cost without adding
setup time. Their identical-length factor two and lower bounds for
arbitrary lengths concern that serial-batch model.

Im and Moseley~\cite{im2013batch} study capacitated batches of requests
for the same page or job type. A request completes after receiving all
pieces of its page. They give flow-time guarantees with resource
augmentation. Their schedules can allocate processing across page
pieces; a TCP-BP launch commits to its admitted set without preemption.
Our comparison pays both launch fees and flow, with the same processor
available to the algorithm and its offline comparator.

Christou et al.~\cite{lookahead2026} study a single server with capacity
\(C\), common batch duration \(s\), and total flow, including adversarial
and Poisson arrivals and imperfect lookahead. Their Theorem~2.4
establishes the optimal deterministic factor two when \(C\ge2\).
For \(C=\infty\), common \(p_j=s\), and \(K=0\), their feasible
schedules and total-flow cost coincide with our TCP-BP subcase.
The equal-length fee-free factor two therefore already holds in that
model. Our total-flow theorem includes a positive fee and arbitrary
lengths through the spread \(\rho\); our batch-maximum objective
aggregates flow once per launched batch. The rooted-purchase, mobile,
and ephemeral-facility comparisons additionally charge spatial service.
The lookahead results address extra arrival information, while our
algorithms use arrivals as they occur.

\paragraph{Active time and flow time.}
Davies, Khuller, and Zhang~\cite{davies2022} minimize total flow for
uniform-length jobs with release times and deadlines, subject to a
budget on synchronized batches of bounded size. Their offline dynamic
programs constrain the number of active batches. Albers and van der
Heijden~\cite{active2026} combine active-slot and total-flow costs in
an online model with integer processing lengths and preemption at
integer slot boundaries. The set of jobs processed may change at
those boundaries. TCP-BP fixes membership at launch and forbids a new
launch until its longest member completes. Its service fee is paid
per launch, independently of that occupied duration. These differences
affect both which schedules are feasible and which commitments incur
additional cost.

\paragraph{Lower-bound technique.}
In online bidding~\cite{bidding}, the cost of a prefix of bids is
compared with the threshold that prompted the last bid. Our long-job
sequence produces such a prefix ratio because all its jobs have the
same value of \(p_j-r_j\): offline can batch them at the final release
and give every job the same flow. The prefix records how often the
algorithm launched. A short job released after the offline processor
is free incurs the remaining online occupation on top of that prefix
cost. This additional charge raises the lower bound above four.

\paragraph{Known lengths and learned scheduling.}
Benomar and Perchet~\cite{benomar2024partial} study partial
processing-size predictions in nonclairvoyant scheduling. Here each
length is known on arrival; the uncertainty is which future jobs will
be blocked by a launch. ED-Batch learns operator batching in dynamic
computation graphs~\cite{chen2023edbatch}, and Packrat configures batch
sizes, threads, and model instances for inference
latency~\cite{bhardwaj2025packrat}. Our fixed-fee and spatial-service
models isolate launch commitments; they do not reproduce those
dependency, memory, or parallel-execution models. In particular, our
competitive guarantees concern the stated fee-plus-flow objectives,
not inference throughput, prediction error, or training quality.

\paragraph{Technical inputs and contribution boundary.}
The instantaneous subadditive maximum-wait guarantee of Lu, Ren, and
Liu~\cite{subadditive} and the instantaneous maximum-wait OSD guarantees
of Lu et al.~\cite{osdmax} are inputs, not new theorems of this paper.
The new comparisons enforce processing occupation while merging
purchases or preserving the order of walks. The single-root upper
and lower bounds do not use either network-service input.

\end{document}